\documentclass[a4paper,10pt]{article}
\usepackage[latin1]{inputenc}
\usepackage[T1]{fontenc}

\usepackage{amsmath}
\usepackage{amsfonts}
\usepackage{amssymb}
\usepackage{xcolor}
\usepackage{graphicx}
\usepackage{caption}
\usepackage{ifthen}
\usepackage{pstricks}
\usepackage{url}
\usepackage{hyperref}
\usepackage[french]{babel}

\usepackage{soul}
\definecolor{bleuclair}{rgb}{0.7, 0.7, 1.0}
\definecolor{rosepale}{rgb}{1.0, 0.7, 1.0}

\usepackage{textcomp}
\usepackage{eurosym}
\usepackage{mathrsfs}

\usepackage{fancybox}
\usepackage{shadow}

\usepackage{wrapfig}
\usepackage{fourier}

\usepackage{geometry}
\usepackage{tikz,tkz-tab}

\begin{document}

\newtheorem{theorem}{Th\'eor\`eme}[section]
\newtheorem{proposition}{Proposition}[section]
\newtheorem{propriete}{Propri\'et\'e}[section]
\newtheorem{lemme}{Lemme}[section]
\newtheorem{corollaire}{Corollaire}[section]
\newtheorem{proof}{Preuve}[section]
\newtheorem{definition}{D\'efinition}[section]
\newtheorem{theoreme-definition}{Th\'eor\`eme et d\'efinition}[section]
\newtheorem{remarque}{Remarque}[section]

\renewcommand{\theproof}{}

\newpage

\section*{
\begin{center}
\shabox{Mod\'elisation spatio-temporelle de l'utilisation des masques par le grand public en Guadeloupe et en Martinique.} 
\end{center}
}

\begin{center}
\vfill

{\bf Jonas SYLNEON} 

\vfill

Contact: sylneonjonas@gmail.com

\vfill
\end{center}

\addcontentsline{toc}{section}{R\'esum\'e}

\section*{R\'esum\'e}
\noindent Dans cet article, nous d\'eveloppons un mod\`ele compartimental d\'eterministe incluant l'aspect spatio-temporel et le port du masque par le grand public pour la propagation de l'\'epid\'emie de Covid-19. Ce mod\`ele est bas\'e sur le mod\`ele SIR de Kermack et McKendrick et comprend des termes de mobilit\'e qui mod\'elisent le d\'eplacement des individus portant des masques entre la Guadeloupe et la Martinique. L'objectif du mod\`ele est d'analyser les effets de la mobilit\'e et du port du masque par le grand public sur la propagation de l'\'epid\'emie de Covid-19 dans ces deux r\'egions. Des simulations num\'eriques de ce mod\`ele montrent l'importance du port du masque par le grand public, m\^eme peu inefficaces, et indiquen que le mod\`ele est capable de simuler qualitativement les tendances de propagation de l'\'epid\'emie de Covid-19. Ce travail est une partie de ma th\`ese de doctorat dont le sujet est intitul\'e \og mod\'elisation stochastique de Covid-19 en milieu insulaire avec prise en compte de l'aspect spatio-temporel \fg.

\section*{Mots cl\'es}
\noindent Mod\`ele spatio-temporel, population, mobilit\'e, restriction, propagation, \'epid\'emie, Covid-19, masque, efficacit\'e, intervention non pharmaceutique.

\addcontentsline{toc}{section}{Abstract}

\section*{Abstract}
\noindent In this article, we develop a deterministic compartmental model including the spatio-temporal aspect and the wearing of masks by the general public for the spread of the Covid-19 epidemic. This model is based on the SIR model of Kermack and McKendrick and includes mobility terms that model the movement of individuals wearing masks between Guadeloupe and Martinique. The objective of the model is to analyze the effects of mobility and the wearing of masks by the general public on the spread of the Covid-19 epidemic in these two regions. Numerical simulations of this model show the importance of wearing masks by the general public, even if they are not very inefficient, and indicate that the model is capable of qualitatively simulating the propagation trends of the Covid-19 epidemic. This work is part of my doctoral thesis, the subject of which is entitled \og stochastic modeling of Covid-19 in an island environment taking into account the spatio-temporal aspect \fg.

\section*{Key words}
\noindent Spatio-temporal model, population, mobility, restriction, spread, epidemic, Covid-19, mask, efficiency, intervention
non-pharmaceutical.

\section{Introduction}
\noindent La maladie Coronavirus 2019, connue sous le nom de Covid-19, caus\'ee par le nouveau coronavirus 2 du syndrome respiratoire aigu s\'ev\`ere (SRAS-CoV2), a \'et\'e signal\'ee pour la premi\`ere fois dans la ville de Wuhan, province du Hubei, en Chine, fin d\'ecembre 2019. L'organisation mondiale de la sant\'e (OMS) a d\'eclar\'e que plus de $80 \ 000$ cas confirm\'es avaient \'et\'e signal\'es dans le monde au 28 f\'evrier 2020 \cite{WHO R-38 (2020)}. Le 11 mars 2020, l'OMS a d\'eclar\'e l'\'epid\'emie de la maladie Covid-19 pand\'emie \cite{WHO D-G (2020)}. La maladie s'est rapidement propag\'ee dans le monde avec $11 \ 937 \ 659$ cas au 7 juillet 2020 \cite{Gayawan and al. (2020)}. Face \`a cette pand\'emie, de nombreux chercheurs d\'eveloppent, testent et ajustent des mod\`eles pour simuler la propagation de cette maladie infectieuse dans le but de mieux la comprendre et d'optimiser les interventions visant \`a la contr\^oler \cite{Arandiga and al. (2020), Peixoto and al. (2020), Arenas and al. (2020)}. \\

\noindent Des chercheurs ont pr\'esent\'e une approche ax\'ee sur les donn\'ees pour la mod\'elisation et la pr\'evision des \'epid\'emies de Covid-19 pouvant \^etre utilis\'ee par les d\'ecideurs pour contr\^oler l'\'epid\'emie gr\^ace \`a des interventions non pharmaceutiques (INP) \cite{Hasan and al. (2020), Eikenberry and al. (2020)}. En France, pour limiter la diffusion de l'\'epid\'emie dans la population, le haut conseil de la sant\'e publique (HCSP) d\'efinit les principales interventions non pharmaceutiques\footnote{Celles qui s'appliquent lorsque ni m\'edicament ni vaccination efficaces ne sont disponibles ou n'existent.}. \\

\noindent La plupart des mod\`eles math\'ematiques de Covid-19 peuvent \^etre globalement divis\'es en mod\`eles de type SIR (Kermack-McKendrick) bas\'es sur la population \cite{Arandiga and al. (2020), Eikenberry and al. (2020), Tang and al. (2020), Li and al. (2020)} ou mod\`eles bas\'es sur des agents \cite{Ferguson and al. (2020), Wilder and al. (2020), Ruiz and Koutronas (2020)}, dans laquelle les individus interagissent g\'en\'eralement sur une structure de r\'eseau et \'echangent l'infection de mani\`ere stochastique. \\

\noindent Dans ce travail, nous proposons un mod\`ele compartimental d\'eterministe avec prise en compte de l'aspect spatial pour \'evaluer l'impact d'une strat\'egie de l'utilisation des masques par le grand public dans les r\'egions de la Guadeloupe et de la Martinique. Ce mod\`ele s'inspire du mod\`ele de Eikenberry \cite{Eikenberry and al. (2020)} et celui de Arandiga \cite{Arandiga and al. (2020)}.

\section{Mod\`eles math\'ematiques}

\subsection{Le mod\`ele SIR avec prise en compte de l'aspect spatial}\label{sir_s}
\noindent Nous consid\'erons le mod\`ele de Arandiga \cite{Arandiga and al. (2020)} sur $2$ zones spatiales: Guadeloupe et Martinique. Nous consid\'erons alors, pour chaque zone $j\in \llbracket 1; 2 \rrbracket$, les compartiments $S_j$, $I_j$ et $R_j$ correspondant respectivement aux susceptibles, infect\'es et retir\'es de la zone $j$. Pour tout $(j, k)\in \llbracket 1; 2 \rrbracket^2$ tel que $k\neq j$, nous notons $n_{jk}$ les flux de mobilit\'e des individus se d\'epla\c{c}ant de la zone $j$ \`a la zone $k$ par unit\'e de temps. Nous consid\'erons que les individus infect\'es symptomatiques et asymptomatiques appartiennent au m\^eme compartiment $I_j$ et ils sont tous infectieux. Nous utilisons l'hypoth\`ese que les individus pr\'esentant des sympt\^omes se sentent malades et restent \`a la maison \cite{Arandiga and al. (2020)}. Nous supposons ainsi que seule une fraction $\eta$ des individus infect\'es se d\'eplace de la Guadeloupe \`a la Martinique et r\'eciproquement. Compte tenu de la dur\'ee du trajet relativement courte entre ces deux r\'egions, nous faisons l'hypoth\`ese que l'infection n'a pas lieu au cours du voyage d'un individus mais seulement dans la r\'egion o\`u celui-ci se trouve \`a un instant $t$ donn\'e. La population totale de ces deux r\'egions est suppos\'ee constante, elle peut varier dans chacune de ces deux r\'egions mais reste globalement constante durant toute la dur\'ee de l'\'epid\'emie\footnote{Nous note cependant que cela est inexact car des naissances et des morts non li\'ees \`a la maladie sont survenues durant cette p\'eriode ainsi que des voyages et des retours notamment vers la m\'etropole.}. Nous supposons que les \'echanges entre ces deux r\'egions sont sym\'etriques. \\

\noindent Il y a $n_{jk}$ individus se d\'epla\c{c}ant de la zone $j$ \`a la zone $k$ par unit\'e de temps, les compartiments $S_j$ et $R_j$ correspondent alors aux compartiments $S_k$ et $R_k$. Puis, le m\^eme nombre d'individus retournent dans la zone $j$ mais certains d'entre eux ont \'et\'e infect\'es pendant leur d\'eplacement dans la zone $k$. \\
Dans ce mod\`ele, $n_{jj}$ correspond aux flux de mobilit\'e des individus se d\'epla\c{c}ant au sein de la zone $j$. \\

\noindent Pour tout $k\in \llbracket 1; 2 \rrbracket$, nous notons $\lambda_k$ la probabilit\'e d'infection des individus de la zone $k$ par unit\'e de temps et $\tau_k$ la dur\'ee du temps qu'un individu susceptible reste dans la zone $k$. Dans la pratique, il existe $(j,k)\in \llbracket 1; 2 \rrbracket^2$ tel que $\tau_k\neq\tau_j$ mais nous supposons que, pour tout $(j,k)\in \llbracket 1; 2 \rrbracket^2$, $\tau_j=\tau_k=\tau$, avec $\tau\in\mathbb{R}^\ast_+$ fix\'e. \\
Du fait que $n_{jk}\dfrac{S_j(t)}{S_j(t)+\eta I_j(t)+R_j(t)}$ individus susceptibles se d\'epla\c{c}ant dans la zone $k$ aient \'et\'e expos\'es \`a la maladie pendant leur d\'eplacement, alors le nombre d'individus infect\'es dans la zone $k$ est:
\begin{eqnarray}\label{(b)}
\tau\lambda_kn_{jk}\dfrac{S_j(t)}{N_j^\ast(t)} ,
\end{eqnarray}
o\`u $N_j^\ast(t)=S_j(t)+\eta I_j(t)+R_j(t)$. Nous avons laiss\'e $\tau$ comme un param\`etre libre dans nos ajustements aux donn\'ees de mobilit\'e entre ces deux r\'egions mais dans \cite{Arandiga and al. (2020)}, Arandiga and al. ont pris $\tau=\dfrac{1}{2}$.  \\ 
La probabilit\'e qu'un susceptible devienne infect\'e dans la zone $j$ est:
$$
\lambda_j=\beta_j\dfrac{I_j(t)}{N_j(t)} ,
$$
o\`u $N_j(t)=S_j(t)+I_j(t)+R_j(t)$. \\
Donc, pour chaque zone $j$, le syst\`eme d'\'equations du mod\`ele SIR sans mobilit\'e s'\'ecrit:
$$
\left\{
\begin{array}{ll}
\dfrac{dS_j}{dt}=-\lambda_j S_j(t) \medskip \\
\dfrac{dI_j}{dt}=\lambda_j S_j(t)-\gamma I_j(t) \medskip \\
\dfrac{dR_j}{dt}=\gamma I_j(t) 
\end{array}
\right.
$$
Les individus susceptibles quittant la zone $j$ par unit\'e de temps sont au nombre de:
$$
\sum_{k=1}^2n_{jk}\dfrac{S_j(t)}{N_j^\ast(t)} .
$$
Tandis que ceux entrant dans la zone $j$ pendant cette m\^eme unit\'e de temps sont au nombre de:
$$
\sum_{k=1}^2n_{kj}\dfrac{S_k(t)}{N_k^\ast(t)} .
$$
Ainsi, par unit\'e de temps, le flux d'individus susceptibles dans la zone $j$ est:
$$
\forall t \in\mathbb{R}_+, \ \varphi_1(t)\sum_{k=1}^2\left( n_{jk}\dfrac{S_j(t)}{N_j^\ast(t)}-n_{kj}\dfrac{S_k(t)}{N_k^\ast(t)}\right) ,
$$
avec $\varphi_1$ la fonction d\'efinie par:
$$
\varphi_1(t)=
\left\{
\begin{array}{ll}
-2 & \mbox{si } t-\lfloor t\rfloor \in[0; \tau[ \\
0  & \mbox{si } t-\lfloor t\rfloor\in[\tau; 1[ \\
\end{array}
\right.
$$
o\`u $\lfloor t \rfloor$ est le plus grand entier inf\'erieur ou \'egal \`a $t$. \\
De la m\^eme mani\`ere, en retour mais en soustrayant le nombre d'individus infect\'es lors de leur d\'eplacement dans la zone $k$, donn\'e dans l'\'equation (\ref{(b)}), le flux d'individus susceptibles dans la zone $j$ est donn\'e par:
$$
\forall t \in\mathbb{R}_+, \ \varphi_2(t)\sum_{k=1}^2\left[n_{jk}\left(\dfrac{S_j(t)}{N_j^\ast(t)}-\tau\dfrac{S_j(t)}{N_j^\ast(t)}\lambda_k\right)- n_{kj}\left(\dfrac{S_k(t)}{N_k^\ast(t)}-\tau\dfrac{S_k(t)}{N_k^\ast(t)}\lambda_j\right)\right] ,
$$
o\`u $\varphi_2$ la fonction d\'efinie par:
$$
\varphi_2(t)=
\left\{
\begin{array}{ll}
0 & \mbox{si } t-\lfloor t\rfloor \in[0; \tau[ \\
2  & \mbox{si } t-\lfloor t\rfloor\in[\tau; 1[ \\
\end{array}
\right.
$$
Par cons\'equent, l'\'equation d'\'evolution des individus susceptibles de la zone $j$ est de la forme:
$$
\begin{array}{rl}
\dfrac{dS_j}{dt} &=-\lambda_jS_j(t)  \medskip \\
&+\varphi_1(t)\displaystyle\sum_{k=1}^2\left( n_{jk}\dfrac{S_j(t)}{N_j^\ast(t)}-n_{kj}\dfrac{S_k(t)}{N_k^\ast(t)}\right) \medskip \\
&+\varphi_2(t)\displaystyle\sum_{k=1}^2\left[n_{jk}\left(\dfrac{S_j(t)}{N_j^\ast(t)}-\tau\dfrac{S_j(t)}{N_j^\ast(t)}\lambda_k\right)- n_{kj}\left(\dfrac{S_k(t)}{N_k^\ast(t)}-\tau\dfrac{S_k(t)}{N_k^\ast(t)}\lambda_j\right)\right] . 
\end{array}
$$
En tenant compte du fait que les termes d\'ecrivant la mobilit\'e des individus du compartiment $I_j$ soient multipli\'es par $\eta$, par un raisonnement analogue, l'\'evolution des individus de ce compartiment est donn\'ee par:
$$
\begin{array}{rl}
\dfrac{dI_j}{dt} &=\lambda_jS_j(t)-\gamma I_j(t) \medskip  \\
&+\eta\varphi_1(t)\displaystyle\sum_{k=1}^2\left( n_{jk}\dfrac{I_j(t)}{N_j^\ast(t)}-n_{kj}\dfrac{I_k(t)}{N_k^\ast(t)}\right) \medskip \\
&+\eta\varphi_2(t)\displaystyle\sum_{k=1}^2\left( n_{jk}\dfrac{I_j(t)}{N_j^\ast(t)}-n_{kj}\dfrac{I_k(t)}{N_k^\ast(t)}\right) \medskip \\
&+\tau\varphi_2(t)\displaystyle\sum_{k=1}^2\left(\dfrac{S_j(t)}{N_j^\ast(t)}\lambda_k+\dfrac{S_k(t)}{N_k^\ast(t)}\lambda_j\right). \\
\end{array}
$$
En utilisant l'hypoth\`ese que les individus du compartiment $R_j$ ne peuvent pas \^etre infect\'es ni infecter les autres, de fa\c{c}on analogue, pour ce compartiment nous avons:
$$
\begin{array}{rl}
\dfrac{dR_j}{dt} &=\gamma I_j(t) \medskip \\
&+\varphi_1(t)\displaystyle\sum_{k=1}^2\left( n_{jk}\dfrac{R_j(t)}{N_j^\ast(t)}-n_{kj}\dfrac{R_k(t)}{N_k^\ast(t)}\right) \medskip \\
&+\varphi_2(t)\displaystyle\sum_{k=1}^2\left( n_{jk}\dfrac{R_j(t)}{N_j^\ast(t)}-n_{kj}\dfrac{R_k(t)}{N_k^\ast(t)}\right) . \\
\end{array}
$$
Finalement, le mod\`ele SIR avec prise en compte de l'aspect spatial est d\'ecrit par le syst\`eme d'\'equations diff\'erentielles ordinaires suivant:
\begin{eqnarray}\label{d}
\left\{
\begin{array}{rl}
\dfrac{dS_j}{dt} &=-\lambda_jS_j(t)
+\varphi(t)\displaystyle\sum_{k=1}^2\left( n_{jk}\dfrac{S_j(t)}{N_j^\ast(t)}-n_{kj}\dfrac{S_k(t)}{N_k^\ast(t)}\right)
-\tau\varphi_2(t)\displaystyle\sum_{k=1}^2\left(n_{jk}\dfrac{S_j(t)}{N_j^\ast(t)}\lambda_k-n_{kj}\dfrac{S_k(t)}{N_k^\ast(t)}\lambda_j\right) \medskip \\ 
\dfrac{dI_j}{dt} &=\lambda_jS_j(t)-\gamma I_j(t)
+\eta\varphi(t)\displaystyle\sum_{k=1}^2\left( n_{jk}\dfrac{I_j(t)}{N_j^\ast(t)}-n_{kj}\dfrac{I_k(t)}{N_k^\ast(t)}\right) 
+\tau\varphi_2(t)\displaystyle\sum_{k=1}^2\left(\dfrac{S_j(t)}{N_j^\ast(t)}\lambda_k+\dfrac{S_k(t)}{N_k^\ast(t)}\lambda_j\right) \medskip \\
\dfrac{dR_j}{dt} &=\gamma I_j(t)
+\varphi(t)\displaystyle\sum_{k=1}^2\left( n_{jk}\dfrac{R_j(t)}{N_j^\ast(t)}-n_{kj}\dfrac{R_k(t)}{N_k^\ast(t)}\right)  \\
\end{array}
\right.
\end{eqnarray}
o\`u $\varphi=\varphi_1+\varphi_2$.  \\

\noindent Nous utiliserons ce mod\`ele pour les taux de transmission r\'esultant de l'utilisation des masques par le grand public dans le mod\`ele propos\'e \`a la section \ref{sir_s1}.  \\

\noindent Pour tout $j\in \llbracket 1; 2 \rrbracket$ et pour tout $t\in\mathbb{R}_+$, nous avons:
$$
\dfrac{dS_j}{dt}+\dfrac{dI_j}{dt}+\dfrac{dR_j}{dt}=\varphi(t)\sum_{k=1}^2\left(n_{jk}-n_{kj}\right) .
$$
C'est-\`a-dire, pour chaque zone $j\in \llbracket 1; 2 \rrbracket$, pour tout temps $t\in\mathbb{R}_+$:
\begin{eqnarray}\label{(c)}
N_j'(t)=\varphi(t)\sum_{k=1}^2( n_{jk}-n_{kj}) .
\end{eqnarray}
Donc, les individus reviennent dans leur zone de r\'esidence apr\`es leur d\'eplacement dans une zone. \\
La solution de l'\'equation (\ref{(c)}) est:
$$
N_j(T)=N_j(0)+\sum_{k=1}^2\left(n_{jk}-n_{kj}\right)\int_{0}^T\varphi(t)dt .
$$
Puisque $\varphi$ est 1-p\'eriodique et $\displaystyle\int_{0}^1\varphi(t)dt=0$ alors, pour tout $T\in\mathbb{N}$, nous avons $N_j(T)=N_j(0)$. \\

\noindent Selon le mod\`ele \cite{Arandiga and al. (2020)}, pour tout $n\in\mathbb{N}$:
$$
N_j\left(n+\tau\right)=N_j(n)+\sum_{k=1}^2\left(n_{jk}-n_{kj}\right)\int_{n}^{n+\tau}\varphi(t)dt=N_j(0)+\sum_{k=1}^z\left(n_{jk}-n_{kj}\right) .
$$
Autrement dit, le mod\`ele indique que les individus passent $\tau$ journ\'ee dans leur zone de r\'esidence et $1-\tau$ journ\'ee dans leur zone de destination.

\subsection{Pr\'esentation du mod\`ele propos\'e}\label{sir_s1}
\noindent Nous supposons qu'une partie de la population totale porte des masques avec un taux d'efficacit\'e int\'erieur (protection contre la maladie) $\varepsilon_i$ et un taux d'efficacit\'e ext\'erieur (protection contre la transmission de la maladie) $\varepsilon_e$. Nous notons $S$, $I$, $R$ les compartiments des individu ne portant pas des masques en public et $\overline{S}$, $\overline{I}$, $\overline{R}$ ceux des individus masqu\'es en public.   \\
Sur la base de ces hypoth\`eses et notations, le mod\`ele SIR avec prise en compte de l'aspect spatial et des masques par le grand public est d\'ecrit par le syst\`eme d'\'equations diff\'erentielles ordinaires non lin\'eaires suivant: 
\begin{eqnarray}\label{mp1}
\left\{
\begin{array}{rl}
\dfrac{dS_j}{dt} &=-\lambda_jS_j(t)
+\varphi(t)\displaystyle\sum_{k=1}^2\left( n_{jk}\dfrac{S_j(t)}{N_j^\ast(t)}-n_{kj}\dfrac{S_k(t)}{N_k^\ast(t)}\right)
-\tau\varphi_2(t)\displaystyle\sum_{k=1}^2\left(n_{jk}\dfrac{S_j(t)}{N_j^\ast(t)}\lambda_k-n_{kj}\dfrac{S_k(t)}{N_k^\ast(t)}\lambda_j\right) \medskip \\ 
&+\left(1-\varepsilon_e\right)\left[-\overline{\lambda}_jS_j(t)
+\varphi(t)\displaystyle\sum_{k=1}^2\left( n_{jk}\dfrac{S_j(t)}{N_j^\ast(t)}-n_{kj}\dfrac{S_k(t)}{N_k^\ast(t)}\right)
-\tau\varphi_2(t)\displaystyle\sum_{k=1}^2\left(n_{jk}\dfrac{S_j(t)}{N_j^\ast(t)}\overline{\lambda}_k-n_{kj}\dfrac{S_k(t)}{N_k^\ast(t)}\overline{\lambda}_j\right)\right] \medskip \\
\dfrac{dI_j}{dt} &=\lambda_jS_j(t)-\gamma I_j(t)
+\eta\varphi(t)\displaystyle\sum_{k=1}^2\left( n_{jk}\dfrac{I_j(t)}{N_j^\ast(t)}-n_{kj}\dfrac{I_k(t)}{N_k^\ast(t)}\right)
+\tau\varphi_2(t)\displaystyle\sum_{k=1}^2\left(\dfrac{S_j(t)}{N_j^\ast(t)}\lambda_k+\dfrac{S_k(t)}{N_k^\ast(t)}\lambda_j\right) \medskip \\
\dfrac{dR_j}{dt} &=\gamma I_j(t)
+\varphi(t)\displaystyle\sum_{k=1}^2\left( n_{jk}\dfrac{R_j(t)}{N_j^\ast(t)}-n_{kj}\dfrac{R_k(t)}{N_k^\ast(t)}\right) \medskip \\
\dfrac{d\overline{S}_j}{dt} &=(1-\varepsilon_i)\left[-\lambda_j\overline{S}_j(t)
+\varphi(t)\displaystyle\sum_{k=1}^2\left( n_{jk}\dfrac{\overline{S}_j(t)}{N_j^\ast(t)}-n_{kj}\dfrac{\overline{S}_k(t)}{N_k^\ast(t)}\right)
-\tau\varphi_2(t)\displaystyle\sum_{k=1}^2\left(n_{jk}\dfrac{\overline{S}_j(t)}{N_j^\ast(t)}\lambda_k-n_{kj}\dfrac{\overline{S}_k(t)}{N_k^\ast(t)}\lambda_j\right)\right] \medskip \\ 
&+(1-\varepsilon_i)(1-\varepsilon_e)\left[-\overline{\lambda}_j\overline{S}_j(t)
+\varphi(t)\displaystyle\sum_{k=1}^2\left( n_{jk}\dfrac{\overline{S}_j(t)}{N_j^\ast(t)}-n_{kj}\dfrac{\overline{S}_k(t)}{N_k^\ast(t)}\right)
-\tau\varphi_2(t)\displaystyle\sum_{k=1}^2\left(n_{jk}\dfrac{\overline{S}_j(t)}{N_j^\ast(t)}\overline{\lambda}_k-n_{kj}\dfrac{\overline{S}_k(t)}{N_k^\ast(t)}\overline{\lambda}_j\right)\right] \medskip \\
\dfrac{d\overline{I}_j}{dt} &=\lambda_j\overline{S}_j(t)-\gamma \overline{I}_j(t)
+\eta\varphi(t)\displaystyle\sum_{k=1}^2\left( n_{jk}\dfrac{\overline{I}_j(t)}{N_j^\ast(t)}-n_{kj}\dfrac{\overline{I}_k(t)}{N_k^\ast(t)}\right)
+\tau\varphi_2(t)\displaystyle\sum_{k=1}^2\left(\dfrac{\overline{S}_j(t)}{N_j^\ast(t)}\lambda_k+\dfrac{\overline{S}_k(t)}{N_k^\ast(t)}\lambda_j\right) \medskip \\
\dfrac{d\overline{R}_j}{dt} &=\gamma \overline{I}_j(t)
+\varphi(t)\displaystyle\sum_{k=1}^2\left( n_{jk}\dfrac{\overline{R}_j(t)}{N_j^\ast(t)}-n_{kj}\dfrac{\overline{R}_k(t)}{N_k^\ast(t)}\right) \medskip \\
\end{array}
\right.
\end{eqnarray}
o\`u $\lambda_j=\beta_j\dfrac{I_j(t)}{N_j(t)}$, $\overline{\lambda}_j=\beta_j\dfrac{\overline{I}_j(t)}{N_j(t)}$ avec, pour tout $(j,t)\in \llbracket 1; 2 \rrbracket \times \mathbb{R}_+$,
$$
N_j(t)=S_j(t)+I_j(t)+R_j(t)+\overline{S}_j(t)+\overline{I}_j(t)+\overline{R}_j(t) ,
$$
et
$$
N_j^\ast(t)=S_j(t)+\eta I_j(t)+R_j(t)+\overline{S}_j(t)+\eta\overline{I}_j(t)+\overline{R}_j(t) .
$$
Pour tout $(j,k)\in \llbracket 1; 2 \rrbracket^2$ tel que $k\neq j$, on pose $n_{jk}=n_{GM}$ et $n_{kj}=n_{MG}$, o\`u G d\'esigne la Guadeloupe et M la Martinique.  \\
Alors, le syst\`eme d'\'equations diff\'erentielles (\ref{mp1}) peut se r\'e\'ecrire de la fa\c{c}on suivante:
\begin{eqnarray}\label{mp2}
\left\{
\begin{array}{rl}
\dfrac{dS_G}{dt} &=-\lambda_GS_G(t)
+\varphi(t)\left( n_{GM}\dfrac{S_G(t)}{N_G^\ast(t)}-n_{MG}\dfrac{S_M(t)}{N_M^\ast(t)}\right)
-\tau\varphi_2(t)\left(n_{GM}\dfrac{S_G(t)}{N_G^\ast(t)}\lambda_M-n_{MG}\dfrac{S_M(t)}{N_M^\ast(t)}\lambda_G\right) \medskip \\ 
&+\left(1-\varepsilon_e\right)\left[-\overline{\lambda}_GS_G(t) 
+\varphi(t)\left( n_{GM}\dfrac{S_G(t)}{N_G^\ast(t)}-n_{MG}\dfrac{S_M(t)}{N_M^\ast(t)}\right)
-\tau\varphi_2(t)\left(n_{GM}\dfrac{S_G(t)}{N_G^\ast(t)}\overline{\lambda}_M-n_{MG}\dfrac{S_M(t)}{N_M^\ast(t)}\overline{\lambda}_G\right)\right] \medskip \\
\dfrac{dI_G}{dt} &=\lambda_GS_G(t)-\gamma I_G(t)
+\eta\varphi(t)\left( n_{GM}\dfrac{I_G(t)}{N_G^\ast(t)}-n_{MG}\dfrac{I_M(t)}{N_M^\ast(t)}\right)
+\tau\varphi_2(t)\left(\dfrac{S_G(t)}{N_G^\ast(t)}\lambda_M+\dfrac{S_M(t)}{N_M^\ast(t)}\lambda_G\right) \medskip \\
\dfrac{dR_G}{dt} &=\gamma I_G(t)
+\varphi(t)\left( n_{GM}\dfrac{R_G(t)}{N_G^\ast(t)}-n_{MG}\dfrac{R_M(t)}{N_M^\ast(t)}\right) \medskip \\
\dfrac{d\overline{S}_G}{dt} &=(1-\varepsilon_i)\left[-\lambda_G\overline{S}_G(t)
+\varphi(t)\left( n_{GM}\dfrac{\overline{S}_G(t)}{N_G^\ast(t)}-n_{MG}\dfrac{\overline{S}_M(t)}{N_M^\ast(t)}\right)
-\tau\varphi_2(t)\left(n_{GM}\dfrac{\overline{S}_G(t)}{N_G^\ast(t)}\lambda_M-n_{MG}\dfrac{\overline{S}_M(t)}{N_M^\ast(t)}\lambda_G\right)\right] \medskip \\ 
&+(1-\varepsilon_i)(1-\varepsilon_e)\left[-\overline{\lambda}_G\overline{S}_G(t)
+\varphi(t)\left( n_{GM}\dfrac{\overline{S}_G(t)}{N_G^\ast(t)}-n_{MG}\dfrac{\overline{S}_M(t)}{N_M^\ast(t)}\right)
-\tau\varphi_2(t)\left(n_{GM}\dfrac{\overline{S}_G(t)}{N_G^\ast(t)}\overline{\lambda}_M-n_{MG}\dfrac{\overline{S}_M(t)}{N_M^\ast(t)}\overline{\lambda}_G\right)\right] \medskip \\
\dfrac{d\overline{I}_G}{dt} &=\lambda_G\overline{S}_G(t)-\gamma \overline{I}_G(t)
+\eta\varphi(t)\left( n_{GM}\dfrac{\overline{I}_G(t)}{N_G^\ast(t)}-n_{MG}\dfrac{\overline{I}_M(t)}{N_M^\ast(t)}\right)
+\tau\varphi_2(t)\left(\dfrac{\overline{S}_G(t)}{N_G^\ast(t)}\lambda_M+\dfrac{\overline{S}_M(t)}{N_M^\ast(t)}\lambda_G\right) \\
\dfrac{d\overline{R}_G}{dt} &=\gamma \overline{I}_G(t)
+\varphi(t)\left( n_{GM}\dfrac{\overline{R}_G(t)}{N_G^\ast(t)}-n_{MG}\dfrac{\overline{R}_M(t)}{N_M^\ast(t)}\right) \medskip \\
\dfrac{dS_M}{dt} &=-\lambda_MS_M(t)
+\varphi(t)\left( n_{MG}\dfrac{S_M(t)}{N_M^\ast(t)}-n_{GM}\dfrac{S_G(t)}{N_G^\ast(t)}\right)
-\tau\varphi_2(t)\left(n_{MG}\dfrac{S_M(t)}{N_M^\ast(t)}\lambda_G-n_{GM}\dfrac{S_G(t)}{N_G^\ast(t)}\lambda_M\right) \medskip \\ 
&+\left(1-\varepsilon_e\right)\left[-\overline{\lambda}_MS_M(t) 
+\varphi(t)\left( n_{MG}\dfrac{S_M(t)}{N_M^\ast(t)}-n_{GM}\dfrac{S_G(t)}{N_G^\ast(t)}\right)
-\tau\varphi_2(t)\left(n_{MG}\dfrac{S_M(t)}{N_M^\ast(t)}\overline{\lambda}_G-n_{GM}\dfrac{S_G(t)}{N_G^\ast(t)}\overline{\lambda}_M\right)\right] \medskip \\
\dfrac{dI_M}{dt} &=\lambda_MS_M(t)-\gamma I_M(t)
+\eta\varphi(t)\left( n_{MG}\dfrac{I_M(t)}{N_M^\ast(t)}-n_{GM}\dfrac{I_G(t)}{N_G^\ast(t)}\right)
+\tau\varphi_2(t)\left(\dfrac{S_M(t)}{N_M^\ast(t)}\lambda_G+\dfrac{S_G(t)}{N_G^\ast(t)}\lambda_M\right) \medskip \\
\dfrac{dR_M}{dt} &=\gamma I_M(t)
+\varphi(t)\left( n_{MG}\dfrac{R_M(t)}{N_M^\ast(t)}-n_{GM}\dfrac{R_G(t)}{N_G^\ast(t)}\right) \medskip \\
\dfrac{d\overline{S}_M}{dt} &=(1-\varepsilon_i)\left[-\lambda_M\overline{S}_M(t)
+\varphi(t)\left( n_{MG}\dfrac{\overline{S}_M(t)}{N_M^\ast(t)}-n_{GM}\dfrac{\overline{S}_G(t)}{N_G^\ast(t)}\right)
-\tau\varphi_2(t)\left(n_{MG}\dfrac{\overline{S}_M(t)}{N_M^\ast(t)}\lambda_G-n_{GM}\dfrac{\overline{S}_G(t)}{N_G^\ast(t)}\lambda_M\right)\right] \medskip \\ 
&+(1-\varepsilon_i)(1-\varepsilon_e)\left[-\overline{\lambda}_M\overline{S}_M(t)
+\varphi(t)\left( n_{MG}\dfrac{\overline{S}_M(t)}{N_M^\ast(t)}-n_{GM}\dfrac{\overline{S}_G(t)}{N_G^\ast(t)}\right)
-\tau\varphi_2(t)\left(n_{MG}\dfrac{\overline{S}_M(t)}{N_M^\ast(t)}\overline{\lambda}_G-n_{GM}\dfrac{\overline{S}_G(t)}{N_G^\ast(t)}\overline{\lambda}_M\right)\right] \medskip \\
\dfrac{d\overline{I}_M}{dt} &=\lambda_M\overline{S}_M(t)-\gamma \overline{I}_M(t)
+\eta\varphi(t)\left( n_{MG}\dfrac{\overline{I}_M(t)}{N_M^\ast(t)}-n_{GM}\dfrac{\overline{I}_G(t)}{N_G^\ast(t)}\right)
+\tau\varphi_2(t)\left(\dfrac{\overline{S}_M(t)}{N_M^\ast(t)}\lambda_G+\dfrac{\overline{S}_G(t)}{N_G^\ast(t)}\lambda_M\right) \medskip \\
\dfrac{d\overline{R}_M}{dt} &=\gamma \overline{I}_M(t)
+\varphi(t)\left( n_{MG}\dfrac{\overline{R}_M(t)}{N_M^\ast(t)}-n_{GM}\dfrac{\overline{R}_G(t)}{N_G^\ast(t)}\right) \medskip \\
\end{array}
\right.
\end{eqnarray}
Nous faisons le changement de variables suivant:
$$
\begin{tabular}{llllll} 
$m_1=\dfrac{\beta_G}{N_G}$; & $m_2=\dfrac{\beta_M}{N_M}$; & $m_3=\dfrac{n_{GM}}{N_G^\ast}$; & $m_4=\dfrac{n_{MG}}{N_M^\ast}$; & $m_5=(1-\varepsilon_e)\dfrac{\beta_G}{N_G}$; \medskip  \\
$m_6=(1-\varepsilon_e)\dfrac{\beta_M}{N_M}$; & $m_7=(1-\varepsilon_e)\dfrac{n_{GM}}{N_G^\ast}$; & $m_8=(1-\varepsilon_e)\dfrac{n_{MG}}{N_M^\ast}$; & $m_9=\dfrac{1}{N_G^\ast}\dfrac{\beta_M}{N_M}$; & $m_{10}=\dfrac{1}{N_M^\ast}\dfrac{\beta_G}{N_G}$. \medskip  \\
\end{tabular}
$$
Puis, pour tout $t\in \mathbb{R}_+$, nous posons:
$$
\begin{tabular}{llllll}
$a_1=m_3\varphi(t)$; & $a_2=m_4\varphi(t)$; & $a_3=\tau m_2m_3\varphi_2(t)$; & $a_4=\tau m_1m_4\varphi_2(t)$; & $a_5=m_7\varphi(t)$; \medskip  \\
$a_6=m_8\varphi(t)$; & $a_7=\tau m_7\varphi_2(t)$; & $a_8=\tau m_8\varphi_2(t)$; & $a_9=\eta m_3\varphi(t)$; & $a_{10}=\eta m_4\varphi(t)$;  \medskip  \\
$a_{11}=\tau m_9\varphi_2(t)$; & $a_{12}=\tau m_{10}\varphi_2(t)$. \medskip  \\
\end{tabular}
$$
Par souci de simplicit\'e, nous omettons la d\'ependance temporelle dans les notations. \\
Nous r\'e\'ecrivons ainsi le syst\`eme d'\'equations non-lin\'eaires (\ref{mp2}) par:
\begin{eqnarray}\label{mp3}
\left\{
\begin{array}{lll}
\dfrac{dS_G}{dt}=-m_1I_GS_G+a_1S_G-a_2S_M-a_3S_GI_M+a_4S_MI_G-m_5\overline{I}_GS_G+a_5S_G-a_6S_M-a_7S_G\overline{I}_M+a_8S_M\overline{I}_G \medskip \\
\dfrac{dI_G}{dt}=m_1I_GS_G-\gamma I_G+a_9I_G-a_{10}I_M+a_{11}S_GI_M+a_{12}S_MI_G \medskip  \\
\dfrac{dR_G}{dt}=\gamma I_G+a_1R_G-a_2R_M \medskip \\
\dfrac{d\overline{S}_G}{dt}=-m_1I_G\overline{S}_G+a_1\overline{S}_G-a_2\overline{S}_M-a_3\overline{S}_GI_M+a_4\overline{S}_MI_G-m_5\overline{I}_G\overline{S}_G+a_5\overline{S}_G-a_6\overline{S}_M-a_7\overline{S}_G\overline{I}_M+a_8\overline{S}_M\overline{I}_G \medskip \\
\dfrac{d\overline{I}_G}{dt}=m_1I_G\overline{S}_G-\gamma \overline{I}_G+a_9\overline{I}_G-a_{10}\overline{I}_M+a_{11}\overline{S}_GI_M+a_{12}\overline{S}_MI_G \medskip \\
\dfrac{d\overline{R}_G}{dt}=\gamma \overline{I}_G+a_1\overline{R}_G-a_2\overline{R}_M \medskip \\
\dfrac{dS_M}{dt}=-m_2I_MS_M+a_2S_M-a_1S_G-a_4S_MI_G+a_3S_GI_M-m_6\overline{I}_MS_M+a_6S_M-a_5S_G-a_4S_M\overline{I}_G+a_3S_G\overline{I}_M \medskip \\
\dfrac{dI_M}{dt}=m_2I_MS_M-\gamma I_M+a_{10}I_M-a_9I_G+a_{12}S_MI_G+a_{11}S_GI_M  \medskip \\
\dfrac{dR_M}{dt}=\gamma I_M+a_2R_M-a_1R_G \medskip \\
\dfrac{d\overline{S}_M}{dt}=-m_2I_M\overline{S}_M+a_2\overline{S}_M-a_1\overline{S}_G-a_4\overline{S}_MI_G+a_3\overline{S}_GI_M-m_6\overline{I}_M\overline{S}_M+a_6\overline{S}_M-a_5\overline{S}_G-a_4\overline{S}_M\overline{I}_G+a_3\overline{S}_G\overline{I}_M \medskip \\
\dfrac{d\overline{I}_M}{dt}=m_2I_M\overline{S}_M-\gamma \overline{I}_M+a_{10}\overline{I}_M-a_9\overline{I}_G+a_{12}\overline{S}_MI_G+a_{11}\overline{S}_GI_M  \medskip \\
\dfrac{d\overline{R}_M}{dt}=\gamma \overline{I}_M+a_2\overline{R}_M-a_1\overline{R}_G \medskip \\
\end{array}
\right.
\end{eqnarray}

\subsection{Existence et unicit\'e de solutions du mod\`ele propos\'e}\label{mbp}
\noindent Nous montrons ici que le probl\`eme est math\'ematiquement bien pos\'e, c'est-\`a-dire qu'il existe bien une solution au syst\`eme (\ref{mp2}). Tout s'y ram\`ene par le th\'eor\`eme suivant:

\begin{theorem}\label{th_EI}
Pour toute condition initiale, il existe une unique solution du syst\`eme d'\'equations (\ref{mp2}).
\end{theorem}

\begin{proof}
Pour tout $t\in\mathbb{R}_+$, nous posons:
$$
x(t)=
\left(
\begin{array}{rl}
S_G(t) \\
I_G(t) \\
R_G(t)  \\
\overline{S}_G(t) \\
\overline{I}_G(t) \\
\overline{R}_G(t)  \\
S_M(t) \\
I_M(t) \\
R_M(t)  \\
\overline{S}_M(t) \\
\overline{I}_M(t) \\
\overline{R}_M(t)  \\
\end{array}
\right)
\ \ \mbox{ et } \ \  
f(t, x(t))=
\left(
\begin{array}{rl}
\dfrac{dS_G}{dt} \medskip \\
\dfrac{dI_G}{dt} \medskip \\
\dfrac{dR_G}{dt} \medskip \\
\dfrac{d\overline{S}_G}{dt} \medskip \\
\dfrac{d\overline{I}_G}{dt} \medskip \\
\dfrac{d\overline{R}_G}{dt} \medskip \\
\dfrac{dS_M}{dt} \medskip \\
\dfrac{dI_M}{dt} \medskip \\
\dfrac{dR_M}{dt} \medskip \\
\dfrac{d\overline{S}_M}{dt} \medskip \\
\dfrac{d\overline{I}_M}{dt} \medskip \\
\dfrac{d\overline{R}_M}{dt} \medskip \\
\end{array}
\right) .
$$
Alors, le syst\`eme d'\'equations (\ref{mp2}) peut s'\'ecrire sous la forme:
\begin{eqnarray}\label{edo_E}
x'(t)=f(t, x(t)) .
\end{eqnarray}
Soit, arbitrairement, $(t_0; x_0)\in \mathbb{R}_+\times\mathbb{R}_+^{12}$ la conditions initiale de notre mod\`ele telle que la solution du syst\`eme (\ref{mp2}) v\'erifie cette condition initiale.  \\
Alors, le probl\`eme de Cauchy associ\'e \`a l'\'equation diff\'erentielle (\ref{edo_E}) est d\'efini par:
\begin{eqnarray}\label{pcm}
\left\{
\begin{array}{ll}
x'(t)=f(t, x(t)) \\
x(t_0)=x_0 \ \mbox{ donn\'e} ,  \\
\end{array}
\right.
\end{eqnarray}
o\`u 
$$
f(x(t))=
A(t)
\left(
\begin{array}{cc}
S_G(t)I_G(t) \\
S_M(t)I_M(t) \\
S_G(t)I_M(t) \\
S_M(t)I_G(t) \\
\overline{S}_G(t)\overline{I}_G(t) \\
\overline{S}_G(t)\overline{I}_M(t) \\
S_G(t)\overline{I}_G(t) \\
\overline{S}_G(t)I_G(t) \\
S_G(t)\overline{I}_M(t) \\
\overline{S}_G(t)I_M(t) \\
\overline{S}_M(t)\overline{I}_M(t) \\
S_M(t)\overline{I}_G(t) \\
S_M(t)\overline{I}_M(t) \\
\overline{S}_M(t)I_G(t) \\
\overline{S}_M(t)\overline{I}_G(t) \\
\overline{S}_M(t)I_M(t) \\
\end{array}
\right)
+
B(t)
\left(
\begin{array}{rl}
S_G(t) \\
I_G(t) \\
R_G(t)  \\
\overline{S}_G(t) \\
\overline{I}_G(t) \\
\overline{R}_G(t)  \\
S_M(t) \\
I_M(t) \\
R_M(t)  \\
\overline{S}_M(t) \\
\overline{I}_M(t) \\
\overline{R}_M(t)  \\
\end{array}
\right) , 
$$
avec $A(t)\in\mathsf{M}_{12,16}(\mathbb{R})$ et $B(t)\in\mathsf{M}_{12}(\mathbb{R})$ sont d\'efinies dans l'annexe \ref{A_EMMP}.  \\
Par ailleurs, $(S_G(t), I_G(t), R_G(t), \overline{S}_G(t), \overline{I}_G(t), \overline{R}_G(t), S_M(t), I_M(t), R_M(t), \overline{S}_M(t), \overline{I}_M(t), \overline{R}_M(t))$ est une base de $\mathbb{R}_+^{12}$ et $x(t)$ d\'esigne une matrice-colonne des coordonn\'ees dans cette base. Si, pour tout $t\in\mathbb{R}_+$, on pose:
$$
\Psi(t, x(t))=
A(t)
\left(
\begin{array}{cc}
S_G(t)I_G(t) \\
S_M(t)I_M(t) \\
S_G(t)I_M(t) \\
S_M(t)I_G(t) \\
\overline{S}_G(t)\overline{I}_G(t) \\
\overline{S}_G(t)\overline{I}_M(t) \\
S_G(t)\overline{I}_G(t) \\
\overline{S}_G(t)I_G(t) \\
S_G(t)\overline{I}_M(t) \\
\overline{S}_G(t)I_M(t) \\
\overline{S}_M(t)\overline{I}_M(t) \\
S_M(t)\overline{I}_G(t) \\
S_M(t)\overline{I}_M(t) \\
\overline{S}_M(t)I_G(t) \\
\overline{S}_M(t)\overline{I}_G(t) \\
\overline{S}_M(t)I_M(t) \\
\end{array}
\right) ,
$$
alors $\Psi$ est une forme bilin\'eaire sym\'etrique d\'efinie sur $\mathbb{R}^{12}$ et, pour tout $t\in\mathbb{R}_+$, on a:
$$
\Psi(t, x(t))=
\left(
\begin{array}{cccccccccccc}
~^tx(t)A_1(t)x(t) \\
~^tx(t)A_2(t)x(t)  \\
~^tx(t)A_3(t)x(t) \\
~^tx(t)A_4(t)x(t)  \\
~^tx(t)A_5(t)x(t)  \\
~^tx(t)A_6(t)x(t) \\
~^tx(t)A_7(t)x(t)  \\
~^tx(t)A_8(t)x(t)  \\
~^tx(t)A_9(t)x(t)  \\
~^tx(t)A_{10}(t)x(t) \\
~^tx(t)A_{11}(t)x(t) \\
~^tx(t)A_{12}(t)x(t) \\
\end{array}
\right) ,
$$
o\`u $~^tx(t)$ d\'esigne la transpos\'ee de la matrice-colonne $x(t)$ et, pour chaque $i\in \llbracket 1; 12 \rrbracket$, $A_i(t)\in\mathsf{M}_{12}(\mathbb{R})$ est d\'efinie dans l'annexe \ref{A_EMMP}.  \\
Ainsi, pour tout $t\in\mathbb{R}_+$, on a:
$$
f(t, x(t))=\Psi(t, x(t))+B(t)x(t) ,
$$ 
o\`u 
$$
B(t)= \left(
\begin{array}{cccccccccccc}
b_{1,1} & 0 & 0 & 0 & 0 & 0 & b_{1,7} & 0 & 0 & 0 & 0 & 0  \\
0 & b_{2,2} & 0 & 0 & 0 & 0 & 0 & 0 & 0 & 0 & 0 & 0  \\
0 & \gamma & \varphi(t)\dfrac{n_{GM}}{N^\ast_G(t)} & 0 & 0 & 0 & 0 & 0 & -\varphi(t)\dfrac{n_{MG}}{N^\ast_M(t)} & 0 & 0 & 0  \\
0 & 0 & 0 &  b_{4,4} & 0 & 0 & 0 & 0 & 0 & b_{4,10} & 0 & 0  \\
0 & 0 & 0 & 0 & b_{5,5} & 0 & 0 & 0 & 0 & 0 & b_{5,11} & 0  \\
0 & 0 & 0 & 0 & \gamma & b_{6,6} & 0 & 0 & 0 & 0 & 0 & b_{6,12}  \\
b_{7,1} & 0 & 0 & 0 & 0 & 0 & b_{7,7} & 0 & 0 & 0 & 0 & 0  \\
0 & 0 & 0 & 0 & 0 & 0 & 0 & b_{8,8} & 0 & 0 & 0 & 0   \\
0 & 0 & -\varphi(t)\dfrac{n_{GM}}{N^\ast_G(t)} & 0 & 0 & 0 & 0 & \gamma & \varphi(t)\dfrac{n_{MG}}{N^\ast_M(t)} & 0 & 0 & 0  \\
0 & 0 & 0 & b_{10,4} & 0 & 0 & 0 & 0 & 0 & b_{10,10} & 0 & 0  \\
0 & 0 & 0 & 0 & b_{11,5} & 0 & 0 & 0 & 0 & 0 & b_{11,11} & 0  \\
0 & 0 & 0 & 0 & 0 & b_{12,6} & 0 & 0 & 0 & 0 & \gamma & b_{12,12}  \\
\end{array}
\right)
$$ 
avec
$$
\begin{tabular}{lll}
$b_{1,1}=(2-\varepsilon_e)\varphi(t)\dfrac{n_{GM}}{N_G^\ast(t)}$, & \qquad $b_{7,1}=-(2-\varepsilon_e)\varphi(t)\dfrac{n_{GM}}{N_G^\ast(t)}$, \medskip \\ 
$b_{2,2}=-\gamma+\eta\varphi(t)\dfrac{n_{GM}}{N_G^\ast(t)}$, & \qquad $b_{8,8}=-\gamma+\eta\varphi(t)\dfrac{n_{MG}}{N_M^\ast(t)}$, & \qquad  \medskip \\
$b_{4,4}=(1-\varepsilon_i)(2-\varepsilon_e)\varphi(t)\dfrac{n_{GM}}{N_G^\ast(t)}$, & \qquad $b_{10,4}=-(1-\varepsilon_i)(2-\varepsilon_e)\varphi(t)\dfrac{n_{GM}}{N_G^\ast(t)}$, \medskip \\
$b_{5,5}=-\gamma+\eta\varphi(t)\dfrac{n_{GM}}{N_G^\ast(t)}$, & \qquad $b_{11,5}=-\eta\varphi(t)\dfrac{n_{GM}}{N_G^\ast(t)}$, \medskip \\
$b_{6,6}=\varphi(t)\dfrac{n_{GM}}{N^\ast_G(t)}$, & \qquad $b_{6,12}=-\varphi(t)\dfrac{n_{MG}}{N^\ast_M(t)}$, \medskip \\ 
$b_{1,7}=-(2-\varepsilon_e)\varphi(t)\dfrac{n_{MG}}{N_M^\ast(t)}$, & \qquad $b_{7,7}=(2-\varepsilon_e)\varphi(t)\dfrac{n_{MG}}{N_M^\ast(t)}$, \medskip \\
$b_{4,10}=-(1-\varepsilon_i)(2-\varepsilon_e)\varphi(t)\dfrac{n_{MG}}{N_M^\ast(t)}$, & \qquad $b_{10,10}=(1-\varepsilon_i)(2-\varepsilon_e)\varphi(t)\dfrac{n_{MG}}{N_M^\ast(t)}$, \medskip \\
$b_{5,11}=-\eta\varphi(t)\dfrac{n_{MG}}{N_M^\ast(t)}$, & \qquad $b_{11,11}=-\gamma+\eta\varphi(t)\dfrac{n_{MG}}{N_M^\ast(t)}$, \medskip \\
$b_{12,6}=-\varphi(t)\dfrac{n_{GM}}{N^\ast_G(t)}$, & \qquad $b_{12,12}=\varphi(t)\dfrac{n_{MG}}{N^\ast_M(t)}$. \medskip \\
\end{tabular}
$$
Montrons que la fonction $f$ est localement lipschitzienne par rapport \`a la deuxi\`eme variable. \\
Dans l'espace vectoriel de dimension finie $\mathbb{R}^{12}$, toutes les normes sont \'equivalentes. Nous munissons $\mathbb{R}^{12}$ de la norme $\parallel . \parallel_1$ et l'espace $\mathsf{M}_n(\mathbb{R})$ des matrices carr\'ees de taille $n$ \`a coefficients r\'eels de la norme $||| . |||_1$. \\ 
Soient $T\in\mathbb{R}_+^\ast$ et $t\in[0; T]$. Alors, pour tout $(x; y) \in \mathbb{R}^{12}\times\mathbb{R}^{12}$, nous avons:
$$
\begin{array}{rl}
\parallel \Psi(t, x(t))-\Psi(t, y(t)) \parallel_1 
&= \displaystyle\sum_{i=1}^{12} \left\lvert ~^tx(t)A_i(t)x(t)-~^ty(t)A_i(t)y(t) \right\rvert \medskip \\
&= \displaystyle\sum_{i=1}^{12} \left\lvert ~^tx(t)A_i(t)x(t)-~^tx(t)A_i(t)y(t)+ ~^tx(t)A_i(t)y(t)-~^ty(t)A_i(t)y(t) \right\rvert \medskip \\
& \leqslant \displaystyle\sum_{i=1}^{12} \left\lvert ~^tx(t)A_i(t)x(t)-~^tx(t)A_i(t)y(t)\right\rvert+\displaystyle\sum_{i=1}^{12}\left\lvert ~^tx(t)A_i(t)y(t)-~^ty(t)A_i(t)y(t) \right\rvert \medskip \\
&= \displaystyle\sum_{i=1}^{12} \left\lvert ~^tx(t)A_i(t)\left(x(t)-y(t)\right)\right\rvert+\displaystyle\sum_{i=1}^{12}\left\lvert 
~^t\left(x(t)-y(t)\right)A_i(t)y(t)\right\rvert \medskip \\
& \leqslant \parallel ~^tx(t)A_i(t) \parallel_p.\parallel x(t)-y(t) \parallel_q +\parallel ~^t\left(x(t)-y(t)\right)\parallel_q.\parallel A_i(t)y(t) \parallel_p , \forall i\in\llbracket 1; 12 \rrbracket \medskip \\
& \ \ \mbox{o\`u} \ (p, q)\in[1; +\infty]^2 \ \mbox{tel que} \ \dfrac{1}{p}+\dfrac{1}{q}=1 \ \mbox{avec} \ \dfrac{1}{\infty}=0 \ (\mbox{par l'in\'egalit\'e de H\"older}) \medskip \\
& \ \ \left(\mbox{car pour} \ (p, q)\in[1; +\infty]^2 \ \mbox{ tel que} \ \dfrac{1}{p}+\dfrac{1}{q}=1, \forall x, y\in\mathbb{R}^{12}, \displaystyle\sum_{i=1}^{12}\left\lvert x_iy_i \right\rvert \leqslant \parallel x \parallel_p\parallel y \parallel_q \right) \medskip \\
& \leqslant \parallel ~^tx(t)A_i(t) \parallel_\infty.\parallel x(t)-y(t) \parallel_1 +\parallel ~^t\left(x(t)-y(t)\right)\parallel_1.\parallel A_i(t)y(t) \parallel_\infty, \forall i\in\llbracket 1; 12 \rrbracket \medskip \\
& \ \ \left(\mbox{par passage \`a la limite car:} \ \forall x\in\mathbb{R}^{12}, \displaystyle\lim_{p\rightarrow +\infty}\parallel x \parallel_p=\parallel x \parallel_\infty \ \mbox{et} \ \displaystyle\lim_{q\rightarrow 1}\parallel x \parallel_q=\parallel x \parallel_1 \right)  \medskip \\
&= \parallel ~^tx(t)A_i(t) \parallel_\infty.\parallel x(t)-y(t) \parallel_1 +\parallel x(t)-y(t) \parallel_1.\parallel A_i(t)y(t) \parallel_\infty, \forall i\in\llbracket 1; 12 \rrbracket \medskip \\
& \ \ \left(\mbox{car:} \ \forall x\in\mathbb{R}^{12}, \parallel ~^tx \parallel_1=\parallel x \parallel_1 \right)  \medskip \\
&\leqslant \left(\parallel ~^tx(t)A_i(t) \parallel_1 + \parallel A_i(t)y(t) \parallel_1\right).\parallel x(t)-y(t) \parallel_1, \forall i\in\llbracket 1; 12 \rrbracket \medskip \\
& \ \ \left(\mbox{car:} \ \forall x\in\mathbb{R}^{12}, \parallel x \parallel_\infty \leqslant \parallel x \parallel_1 \right)  \medskip \\
& \leqslant \left(\parallel ~^tx(t) \parallel_1. ||| A_i(t) |||_1 + ||| A_i(t) |||_1.\parallel y(t) \parallel_1\right).\parallel x(t)-y(t) \parallel_1, \forall i\in\llbracket 1; 12 \rrbracket  \medskip \\

& \ \ \left(\mbox{car:} \ \forall A\in\mathsf{M}_{12}(\mathbb{R}), \forall x\in\mathbb{R}^{12}, \parallel Ax \parallel_1 \leqslant ||| A |||_1 \parallel x \parallel_1 \right)  \medskip \\ 

&= \left(\parallel x(t) \parallel_1+\parallel y(t) \parallel_1 \right).||| A_i(t) |||_1.\parallel x(t)-y(t) \parallel_1, \forall i\in\llbracket 1; 12 \rrbracket \medskip \\
&= K(t) \parallel x(t)-y(t) \parallel_1 , \medskip \\
\end{array}
$$
o\`u, pour $t\in[0; T]$, 
$$
K(t)=\left(\parallel x(t) \parallel_1+\parallel y(t) \parallel_1 \right) \displaystyle \min_{1\leqslant i \leqslant 12} ||| A_i(t) |||_1 >0 .
$$
Il s'ensuit que, pour $t\in[0; T]$ et pour tout $(x; y) \in \mathbb{R}^{12}\times\mathbb{R}^{12}$:
$$
\begin{array}{rl}
\parallel f(t, x(t))-f(t, y(t)) \parallel_1 
&= \parallel \Psi(t, x(t))-\Psi(t, y(t))+B(t)(x(t)-y(t)) \parallel_1 \medskip \\
&\leqslant \parallel \Psi(t, x(t))-\Psi(t, y(t))\parallel_1 + \parallel B(t)(x(t)-y(t)) \parallel_1 \medskip \\
&\leqslant \parallel \Psi(t, x(t))-\Psi(t, y(t)) \parallel_1 + ||| B(t) |||_1.\parallel x(t)-y(t) \parallel_1 \medskip \\
&\leqslant K(t) \parallel x(t)-y(t) \parallel_1 + ||| B(t) |||_1.\parallel x(t)-y(t) \parallel_1 \medskip \\
&= L(t) \parallel x(t)-y(t) \parallel_1 , \medskip \\
\end{array}
$$
avec, pour $t\in[0; T]$,
$$
\begin{array}{rl}
L(t) &= K(t) + ||| B(t) |||_1  \medskip \\
&= \left(\parallel x(t) \parallel_1+\parallel y(t) \parallel_1 \right) \displaystyle \min_{1\leqslant i \leqslant 12} ||| A_i(t) |||_1 + ||| B(t) |||_1 > 0  \medskip \\
\end{array}
$$
Comme de plus $L$ est born\'ee, on en d\'eduit que $f$ est localement lipschitzienne  sur $\mathbb{R}_+\times\mathbb{R}^{12}$ par rapport \`a la seconde variable (la variable d'\'etat). \\
Ainsi, par le th\'eor\`eme de Cauchy-Lipschitz, le probl\`eme de Cauchy (\ref{pcm}) admet une unique solution. 
\hfill $\Box$
\end{proof}

\subsection{Ensemble compact positivement invariant du mod\`ele propos\'e}
\noindent Dans la section \ref{mbp}, nous avons montr\'e que le mod\`ele est math\'ematiquement bien pos\'e. Pour s'assurer que le mod\`ele soit \'epidemilogiquement bien pos\'e, nous allons d\'efinir un ensemble compact positivement invariant $\Omega$ qui aura la propri\'et\'e que toute trajectoire du syst\`eme (\ref{mp3}) commen\c{c}ant dans $\Omega$ y reste pour tout temps futur. Le th\'eor\`eme suivant est de mise.

\begin{theorem}
L'ensemble 
$$
\Omega=\{(S_j,I_j,R_j,\overline{S}_j,\overline{I}_j,\overline{R}_j)\in\mathbb{R}_+^6 \mid 0\leq S_j+I_j+R_j+\overline{S}_j+\overline{I}_j+\overline{R}_j\leqslant N_j(0), j \in \{G; M \}\} .
$$
est un compact positivement invariant pour le syst\`eme(\ref{mp3}). 
\end{theorem}
\begin{proof}
Montrons que $\Omega$ est un ensemble positif. Il s'agit alors de montrer que toutes les variables d'\'etat restent positives \`a chaque instant $t\in\mathbb{R}_+$ puisqu'elles repr\'esentent des nombres d'individus. \\
Pour commencer, nous supposons que toutes les conditions initiales sont positives ou strictement positives:
$$
\forall j\in \{G; M \}, \ S_j(0) > 0, \ I_j(0) > 0,\  R_j(0) \geqslant 0, \ \overline{S}_j(0) > 0, \ \overline{I}_j(0) > 0, \ \overline{R}_j(0) \geqslant 0 .
$$
Alors, il existe un certain laps de temps $r_0\in\mathbb{R}_+^\ast$ tel que, pendant le temps $t\in[0; r_0]$, toutes les variables d'\'etat restent positives ou strictement positives jusqu'au moment o\`u l'une d'entre elles s'annule.  \\
Donc, d'apr\`es la premi\`ere \'equation du syst\`eme (\ref{mp3}), en posant $t_0=0$ et $t_1=r_0>0$, pour chaque instant $t\in[t_0; t_1]$, nous avons:
$$
\begin{array}{rl}
\dfrac{dS_G}{dt} &= -m_1I_GS_G+a_1S_G-a_2S_M-a_3S_GI_M+a_4S_MI_G-m_5\overline{I}_GS_G+a_5S_G-a_6S_M-a_7S_G\overline{I}_M+a_8S_M\overline{I}_G \medskip \\
               &\geqslant -m_1I_GS_G+a_1S_G-a_2S_M-a_3S_GI_M-m_5\overline{I}_GS_G+a_5S_G-a_6S_M-a_7S_G\overline{I}_M \medskip \\
               &\mbox{ \ \ \ car, $m_1 > 0$, $m_5 \geqslant 0$ et, pour tout $t\in\mathbb{R}_+$, $a_3 \geqslant 0$, $a_4 \geqslant 0$, $a_7 \geqslant 0$, $a_8 \geqslant 0$.} \medskip \\
               & \geqslant 
\left\{              
\begin{array}{ll}
-m_1I_GS_G-a_2S_M-a_3S_GI_M-m_5\overline{I}_GS_G-a_6S_M-a_7S_G\overline{I}_M & \mbox{ si } \ \varphi(t) >0  \medskip \\
-m_1I_GS_G+a_1S_G-a_3S_GI_M-m_5\overline{I}_GS_G+a_5S_G-a_7S_G\overline{I}_M & \mbox{ si } \ \varphi(t) <0 \medskip \\
\end{array}
\right. \medskip \\
               &= 
\left\{              
\begin{array}{ll}
-S_G(m_1I_G+a_3I_M+m_5\overline{I}_G+a_7\overline{I}_M)-(a_2+a_6)S_M & \mbox{ si } \ \varphi(t) >0  \medskip \\
-S_G(m_1I_G-a_1+a_3I_M+m_5\overline{I}_G-a_5+a_7\overline{I}_M) & \mbox{ si } \ \varphi(t) <0 \medskip \\
\end{array}
\right. \medskip \\ 
              &\geqslant
\left\{              
\begin{array}{ll}
-S_G(m_1I_G+a_3I_M+m_5\overline{I}_G+a_7\overline{I}_M +a_2+a_6) & \mbox{ si } \ \varphi(t) >0  \medskip \\
-S_G(m_1I_G-a_1+a_3I_M+m_5\overline{I}_G-a_5+a_7\overline{I}_M) & \mbox{ si } \ \varphi(t) <0 \medskip \\
\end{array}
\right. \medskip \\ 
& \ \ \ (\mbox{en supposant arbitrairement que $N_G>N_M$ et donc, pour tout $t\in[t_0; t_1]$, $S_G(t)>S_M(t)$}) \medskip \\
              &=
\left\{              
\begin{array}{ll}
-S_G(m_1I_G+a_3I_M+m_5\overline{I}_G+a_7\overline{I}_M +a_2+a_6) & \mbox{ si } \ \varphi(t) >0 \ \mbox{ et } \ \varphi_2(t)=2  \medskip \\
-S_G(m_1I_G-a_1+a_3I_M+m_5\overline{I}_G-a_5+a_7\overline{I}_M) & \mbox{ si } \ \varphi(t) <0 \ \mbox{ et } \ \varphi_2(t)=2 \medskip \\
-S_G(m_1I_G+m_5\overline{I}_G+a_2+a_6) & \mbox{ si } \ \varphi(t) >0 \ \mbox{ et } \ \varphi_2(t) = 0  \medskip \\
-S_G(m_1I_G-a_1+m_5\overline{I}_G-a_5) & \mbox{ si } \ \varphi(t) <0 \ \mbox{ et } \ \varphi_2(t) = 0 \medskip \\
\end{array}
\right. \medskip \\ 
\end{array}
$$
D'o\`u, pour tout $t\in[t_0; t_1]$, nous avons:
$$\left\{              
\begin{array}{ll}
\dfrac{dS_G}{dt}+S_G(m_1I_G+a_3I_M+m_5\overline{I}_G+a_7\overline{I}_M +a_2+a_6) \geqslant 0 & \mbox{ si } \ \varphi(t) >0 \ \mbox{ et } \ \varphi_2(t)=2  \medskip \\
\dfrac{dS_G}{dt}+S_G(m_1I_G-a_1+a_3I_M+m_5\overline{I}_G-a_5+a_7\overline{I}_M) \geqslant 0 & \mbox{ si } \ \varphi(t) <0 \ \mbox{ et } \ \varphi_2(t)=2 \medskip \\
\dfrac{dS_G}{dt}+S_G(m_1I_G+m_5\overline{I}_G+a_2+a_6) \geqslant 0 & \mbox{ si } \ \varphi(t) >0 \ \mbox{ et } \ \varphi_2(t) = 0  \medskip \\
\dfrac{dS_G}{dt}+S_G(m_1I_G-a_1+m_5\overline{I}_G-a_5) \geqslant 0 & \mbox{ si } \ \varphi(t) <0 \ \mbox{ et } \ \varphi_2(t) = 0 \medskip \\
\end{array}
\right. 
$$
Nous notons $I=(I_G, I_M, \overline{I}_G, \overline{I}_M)$ et posons:
$$
F(t, I(t))=\min_{t\in[t_0; t_1]}
\left\{              
\begin{array}{ll}
m_1I_G+a_3I_M+m_5\overline{I}_G+a_7\overline{I}_M +a_2+a_6 & \mbox{ si } \ \varphi(t) >0 \ \mbox{ et } \ \varphi_2(t)=2  \medskip \\
m_1I_G-a_1+a_3I_M+m_5\overline{I}_G-a_5+a_7\overline{I}_M & \mbox{ si } \ \varphi(t) <0 \ \mbox{ et } \ \varphi_2(t)=2 \medskip \\
m_1I_G+m_5\overline{I}_G+a_2+a_6 & \mbox{ si } \ \varphi(t) >0 \ \mbox{ et } \ \varphi_2(t) = 0  \medskip \\
m_1I_G-a_1+m_5\overline{I}_G-a_5 & \mbox{ si } \ \varphi(t) <0 \ \mbox{ et } \ \varphi_2(t) = 0 \medskip \\
\end{array}
\right. \\ 
$$
Alors, pour tout $t\in[t_0; t_1]$, nous avons:
$$
\dfrac{dS_G}{dt}+S_G(t)F(t, I(t)) \geqslant 0 .
$$
Pour tout $t\in[t_0; t_1]$, en multipliant les deux membres de l'in\'egalit\'e par $\exp \left(\displaystyle\int_{t_0}^t F(s, I(s))ds \right)>0$, nous obtenons:
$$
\exp\left(\int_{t_0}^t F(s, I(s))ds \right)\times\dfrac{dS_G(t)}{dt} + F(t, I(t))\exp\left(\displaystyle\int_{t_0}^t F(s,I(s))ds \right)\times S_G(t) \geqslant 0 .
$$
Soit, pour tout $t\in[t_0; t_1]$:
$$
\dfrac{d}{dt}\left[S_G(t)\times \exp\left(\int_{t_0}^t F(s, I(s))ds \right)\right] \geqslant 0 .
$$
Ensuite, pour tout $t\in[t_0; t_1]$, par int\'egration entre $t_0$ et $t$, nous avons:
$$
\int_{t_0}^t \dfrac{d}{ds}\left[S_G(s) \exp\left(\int_{t_0}^t F(s, I(s))ds \right)\right] ds \geqslant 0 .
$$
Soit, pour tout $t\in[t_0; t_1]$:
$$
S_G(t)\exp\left(\int_{t_0}^t F(s, I(s))ds \right) \geqslant S_G(t_0)\exp\left(\int_{t_0}^t F(s, I(s))ds \right) .
$$
Puisque, pour tout $t\in[t_0; t_1]$, $F(t, I(t)) \geqslant 0$, alors nous avons $\displaystyle \int_{t_0}^t F(s, I(s))ds \geqslant 0$.  \\
Puis, pour tout $t\in[t_0; t_1]$, nous avons $\exp\left(\displaystyle\int_{t_0}^t F(s, I(s))ds \right) \geqslant 1$. \\
Nous en d\'eduisons que, pour tout $t\in[t_0; t_1]$:
$$
S_G(t)\geqslant S_G(t_0) \exp\left(-\int_{t_0}^t F(s, I(s))ds \right) .
$$
Donc, pour tout $t\in[t_0; t_1]$, nous avons $S_G(t) > 0$.  \\
D'apr\`es la deuxi\`eme \'equation du syst\`eme (\ref{mp3}), en posant $t_0=0$ et $t_1=r_0>0$, pour tout temps $t\in[t_0; t_1]$, nous avons:
$$
\begin{array}{rl}
\dfrac{dI_G}{dt} &= m_1I_GS_G-\gamma I_G+a_9I_G-a_{10}I_M+a_{11}S_GI_M+a_{12}S_MI_G \medskip \\
               &\geqslant -\gamma I_G+a_9I_G-a_{10}I_M \ \ \mbox{ (car, $m_1 > 0$ et, pour tout $t\in\mathbb{R}_+$, $a_{11} \geqslant 0$, $a_{12} \geqslant 0$)} \medskip \\
               & \geqslant 
\left\{              
\begin{array}{ll}
-\gamma I_G-a_{10}I_M & \mbox{ si } \ \varphi(t) >0  \medskip \\
-\gamma I_G+a_9I_G & \mbox{ si } \ \varphi(t) <0 \medskip \\
\end{array}
\right. \medskip \\
               &= 
\left\{              
\begin{array}{ll}
-(\gamma I_G+a_{10}I_M) & \mbox{ si } \ \varphi(t) >0  \medskip \\
-(\gamma-a_9)I_G & \mbox{ si } \ \varphi(t) <0 \medskip \\
\end{array}
\right. \medskip \\ 
              &\geqslant
\left\{              
\begin{array}{ll}
-(\gamma+a_{10})I_G & \mbox{ si } \ \varphi(t) >0  \medskip \\
-(\gamma-a_9)I_G & \mbox{ si } \ \varphi(t) <0 \medskip \\
\end{array}
\right. \medskip \\ 
& \ \ \ (\mbox{en prenant arbitrairement $I_G(t_0)>I_M(t_0)$ et donc, pour tout $t\in[t_0; t_1]$, $I_G(t)>I_M(t)$}). \medskip \\
\end{array}
$$
D'o\`u, pour tout $t\in[t_0; t_1]$, nous avons:
$$\left\{              
\begin{array}{ll}
\dfrac{dI_G}{dt}+(\gamma+a_{10})I_G \geqslant 0 & \mbox{ si } \ \varphi(t) >0  \medskip \\
\dfrac{dI_G}{dt}+(\gamma-a_9)I_G \geqslant 0 & \mbox{ si } \ \varphi(t) <0 \medskip \\
\end{array}
\right. 
$$
Nous posons:
$$
F(t)=\min_{t\in[t_0; t_1]}
\left\{              
\begin{array}{ll}
\gamma+a_{10} & \mbox{ si } \ \varphi(t) >0  \medskip \\
\gamma-a_9 & \mbox{ si } \ \varphi(t) <0 \medskip \\
\end{array}
\right. \\ 
$$
Alors, pour tout $t\in[t_0; t_1]$, nous avons:
$$
\dfrac{dI_G}{dt}+I_G(t)F(t) \geqslant 0 .
$$
Pour tout $t\in[t_0; t_1]$, en multipliant les deux membres de l'in\'egalit\'e par $\exp \left(\displaystyle\int_{t_0}^t F(s)ds \right)>0$, nous obtenons:
$$
\exp\left(\int_{t_0}^t F(s)ds \right)\times\dfrac{dI_G(t)}{dt} + F(t)\exp\left(\displaystyle\int_{t_0}^t F(s)ds \right)\times I_G(t) \geqslant 0 .
$$
Soit, pour tout $t\in[t_0; t_1]$:
$$
\dfrac{d}{dt}\left[I_G(t)\times \exp\left(\int_{t_0}^t F(s)ds \right)\right] \geqslant 0 .
$$
Ensuite, pour tout $t\in[t_0; t_1]$, par int\'egration entre $t_0$ et $t$, nous avons:
$$
\int_{t_0}^t \dfrac{d}{ds}\left[I_G(s) \exp\left(\int_{t_0}^t F(s)ds \right)\right] ds \geqslant 0 .
$$
Nous en d\'eduisons que, pour tout $t\in[t_0; t_1]$:
$$
I_G(t)\geqslant I_G(t_0) \exp\left(-\int_{t_0}^t F(s)ds \right) .
$$
Donc, pour tout $t\in[t_0; t_1]$, nous avons $I_G(t) > 0$.  \\
D'apr\`es la troisi\`eme \'equation du syst\`eme (\ref{mp3}), en posant $t_0=0$ et $t_1=r_0>0$, pour tout temps $t\in[t_0; t_1]$, nous avons:
$$
\begin{array}{rl}
\dfrac{dR_G}{dt} &= \gamma I_G+a_1R_G-a_2R_M \medskip \\
               &\geqslant a_1R_G-a_2R_M \ \ \mbox{ (car $\gamma > 0$)} \medskip \\
               & \geqslant 
\left\{              
\begin{array}{ll}
-a_2R_M & \mbox{ si } \ \varphi(t) >0  \medskip \\
a_1R_G & \mbox{ si } \ \varphi(t) <0 \medskip \\
\end{array}
\right. \medskip \\
               &\geqslant
\left\{              
\begin{array}{ll}
-a_2R_G & \mbox{ si } \ \varphi(t) >0  \medskip \\
a_1R_G & \mbox{ si } \ \varphi(t) <0 \medskip \\
\end{array}
\right. \medskip \\ 
& \ \ \ (\mbox{en supposant arbitrairement que $N_G>N_M$ et donc, pour tout $t\in[t_0; t_1]$, $R_G(t)>R_M(t)$}). \medskip \\
\end{array}
$$
D'o\`u, pour tout $t\in[t_0; t_1]$, nous avons:
$$\left\{              
\begin{array}{ll}
\dfrac{dR_G}{dt}+a_2R_G \geqslant 0 & \mbox{ si } \ \varphi(t) >0  \medskip \\
\dfrac{dR_G}{dt}-a_1R_G \geqslant 0 & \mbox{ si } \ \varphi(t) <0 \medskip \\
\end{array}
\right. 
$$
Nous posons:
$$
F(t)=\min_{t\in[t_0; t_1]}
\left\{              
\begin{array}{ll}
a_2 & \mbox{ si } \ \varphi(t) >0  \medskip \\
-a_1 & \mbox{ si } \ \varphi(t) <0 \medskip \\
\end{array}
\right. \\ 
$$
Alors, pour tout $t\in[t_0; t_1]$, nous avons:
$$
\dfrac{dR_G}{dt}+R_G(t)F(t) \geqslant 0 .
$$
Pour tout $t\in[t_0; t_1]$, en multipliant les deux membres de l'in\'egalit\'e par $\exp \left(\displaystyle\int_{t_0}^t F(s)ds \right)>0$, nous obtenons:
$$
\exp\left(\int_{t_0}^t F(s)ds \right)\times\dfrac{dR_G(t)}{dt} + F(t)\exp\left(\displaystyle\int_{t_0}^t F(s)ds \right)\times R_G(t) \geqslant 0 .
$$
Soit, pour tout $t\in[t_0; t_1]$:
$$
\dfrac{d}{dt}\left[R_G(t)\times \exp\left(\int_{t_0}^t F(s)ds \right)\right] \geqslant 0 .
$$
Ensuite, pour tout $t\in[t_0; t_1]$, par int\'egration entre $t_0$ et $t$, nous avons:
$$
\int_{t_0}^t \dfrac{d}{ds}\left[R_G(s) \exp\left(\int_{t_0}^t F(s)ds \right)\right] ds \geqslant 0 .
$$
Nous en d\'eduisons que, pour tout $t\in[t_0; t_1]$:
$$
R_G(t)\geqslant R_G(t_0) \exp\left(-\int_{t_0}^t F(s)ds \right) .
$$
Donc, pour tout $t\in[t_0; t_1]$, nous avons $R_G(t) \geqslant 0$.  \\
D'apr\`es la quatri\`eme \'equation du syst\`eme (\ref{mp3}), en posant $t_0=0$ et $t_1=r_0>0$, pour tout temps $t\in[t_0; t_1]$, nous avons:
$$
\begin{array}{rl}
\dfrac{d\overline{S}_G}{dt} &= -m_1I_G\overline{S}_G+a_1\overline{S}_G-a_2\overline{S}_M-a_3\overline{S}_GI_M+a_4\overline{S}_MI_G-m_5\overline{I}_G\overline{S}_G+a_5\overline{S}_G-a_6\overline{S}_M-a_7\overline{S}_G\overline{I}_M+a_8\overline{S}_M\overline{I}_G \medskip \\
               &\geqslant -m_1I_G\overline{S}_G+a_1\overline{S}_G-a_2\overline{S}_M-a_3\overline{S}_GI_M-m_5\overline{I}_G\overline{S}_G+a_5\overline{S}_G-a_6\overline{S}_M-a_7\overline{S}_G\overline{I}_M \medskip \\
               &\mbox{ \ \ \ (car, $m_1 > 0$, $m_5 \geqslant 0$ et, pour tout $t\in\mathbb{R}_+$, $a_3 \geqslant 0$, $a_4 \geqslant 0$, $a_7 \geqslant 0$, $a_8 \geqslant 0$)} \medskip \\
               & \geqslant 
\left\{              
\begin{array}{ll}
-m_1I_G\overline{S}_G-a_2\overline{S}_M-a_3\overline{S}_GI_M-m_5\overline{I}_G\overline{S}_G-a_6\overline{S}_M-a_7\overline{S}_G\overline{I}_M & \mbox{ si } \ \varphi(t) >0  \medskip \\
-m_1I_G\overline{S}_G+a_1\overline{S}_G-a_3\overline{S}_GI_M-m_5\overline{I}_G\overline{S}_G+a_5\overline{S}_G-a_7\overline{S}_G\overline{I}_M & \mbox{ si } \ \varphi(t) <0 \medskip \\
\end{array}
\right. \medskip \\
               &= 
\left\{              
\begin{array}{ll}
-\overline{S}_G(m_1I_G+a_3I_M+m_5\overline{I}_G+a_7\overline{I}_M)-(a_2+a_6)\overline{S}_M  & \mbox{ si } \ \varphi(t) >0  \medskip \\
-\overline{S}_G(m_1I_G-a_1+a_3I_M+m_5\overline{I}_G-a_5+a_7\overline{I}_M) & \mbox{ si } \ \varphi(t) <0 \medskip \\
\end{array}
\right. \medskip \\ 
              &\geqslant
\left\{              
\begin{array}{ll}
-\overline{S}_G(m_1I_G+a_3I_M+m_5\overline{I}_G+a_7\overline{I}_M+a_2+a_6)  & \mbox{ si } \ \varphi(t) >0  \medskip \\
-\overline{S}_G(m_1I_G-a_1+a_3I_M+m_5\overline{I}_G-a_5+a_7\overline{I}_M) & \mbox{ si } \ \varphi(t) <0 \medskip \\
\end{array}
\right. \medskip \\ 
& \ \ \ (\mbox{en supposant arbitrairement que $N_G>N_M$ et donc, pour tout $t\in[t_0; t_1]$, $\overline{S}_G(t)>\overline{S}_M(t)$}) \medskip \\
              &=
\left\{              
\begin{array}{ll}
-\overline{S}_G(m_1I_G+a_3I_M+m_5\overline{I}_G+a_7\overline{I}_M+a_2+a_6) & \mbox{ si } \ \varphi(t) >0 \ \mbox{ et } \ \varphi_2(t)=2  \medskip \\
-\overline{S}_G(m_1I_G-a_1+a_3I_M+m_5\overline{I}_G-a_5+a_7\overline{I}_M) & \mbox{ si } \ \varphi(t) <0 \ \mbox{ et } \ \varphi_2(t)=2 \medskip \\
-\overline{S}_G(m_1I_G+m_5\overline{I}_G+a_2+a_6) & \mbox{ si } \ \varphi(t) >0 \ \mbox{ et } \ \varphi_2(t) = 0  \medskip \\
-\overline{S}_G(m_1I_G-a_1+m_5\overline{I}_G-a_5) & \mbox{ si } \ \varphi(t) <0 \ \mbox{ et } \ \varphi_2(t) = 0 \medskip \\
\end{array}
\right. \medskip \\ 
\end{array}
$$
D'o\`u, pour tout $t\in[t_0; t_1]$, nous avons:
$$\left\{              
\begin{array}{ll}
\dfrac{d\overline{S}_G}{dt}+\overline{S}_G(m_1I_G+a_3I_M+m_5\overline{I}_G+a_7\overline{I}_M+a_2+a_6) \geqslant 0 & \mbox{ si } \ \varphi(t) >0 \ \mbox{ et } \ \varphi_2(t)=2  \medskip \\
\dfrac{d\overline{S}_G}{dt}+\overline{S}_G(m_1I_G-a_1+a_3I_M+m_5\overline{I}_G-a_5+a_7\overline{I}_M) \geqslant 0 & \mbox{ si } \ \varphi(t) <0 \ \mbox{ et } \ \varphi_2(t)=2 \medskip \\
\dfrac{d\overline{S}_G}{dt}+\overline{S}_G(m_1I_G+m_5\overline{I}_G+a_2+a_6) \geqslant 0 & \mbox{ si } \ \varphi(t) >0 \ \mbox{ et } \ \varphi_2(t) = 0  \medskip \\
\dfrac{d\overline{S}_G}{dt}+\overline{S}_G(m_1I_G-a_1+m_5\overline{I}_G-a_5) \geqslant 0 & \mbox{ si } \ \varphi(t) <0 \ \mbox{ et } \ \varphi_2(t) = 0 \medskip \\
\end{array}
\right. 
$$
Nous notons $I=(I_G, I_M, \overline{I}_G, \overline{I}_M)$ et posons:
$$
F(t, I(t))=\min_{t\in[t_0; t_1]}
\left\{              
\begin{array}{ll}
m_1I_G+a_3I_M+m_5\overline{I}_G+a_7\overline{I}_M+a_2+a_6 & \mbox{ si } \ \varphi(t) >0 \ \mbox{ et } \ \varphi_2(t)=2  \medskip \\
m_1I_G-a_1+a_3I_M+m_5\overline{I}_G-a_5+a_7\overline{I}_M & \mbox{ si } \ \varphi(t) <0 \ \mbox{ et } \ \varphi_2(t)=2 \medskip \\
m_1I_G+m_5\overline{I}_G+a_2+a_6 & \mbox{ si } \ \varphi(t) >0 \ \mbox{ et } \ \varphi_2(t) = 0  \medskip \\
m_1I_G-a_1+m_5\overline{I}_G-a_5 & \mbox{ si } \ \varphi(t) <0 \ \mbox{ et } \ \varphi_2(t) = 0 \medskip \\
\end{array}
\right. \\ 
$$
Alors, pour tout $t\in[t_0; t_1]$, nous avons:
$$
\dfrac{d\overline{S}_G}{dt}+\overline{S}_G(t)F(t, I(t)) \geqslant 0 .
$$
Pour tout $t\in[t_0; t_1]$, en multipliant les deux membres de l'in\'egalit\'e par $\exp \left(\displaystyle\int_{t_0}^t F(s, I(s))ds \right)>0$, nous obtenons:
$$
\exp\left(\int_{t_0}^t F(s, I(s))ds \right)\times\dfrac{d\overline{S}_G(t)}{dt} + F(t, I(t))\exp\left(\displaystyle\int_{t_0}^t F(s,I(s))ds \right)\times \overline{S}_G(t) \geqslant 0 .
$$
Soit, pour tout $t\in[t_0; t_1]$:
$$
\dfrac{d}{dt}\left[\overline{S}_G(t)\times \exp\left(\int_{t_0}^t F(s, I(s))ds \right)\right] \geqslant 0 .
$$
Ensuite, pour tout $t\in[t_0; t_1]$, par int\'egration entre $t_0$ et $t$, nous avons:
$$
\int_{t_0}^t \dfrac{d}{ds}\left[\overline{S}_G(s) \exp\left(\int_{t_0}^t F(s, I(s))ds \right)\right] ds \geqslant 0 .
$$
Nous en d\'eduisons que, pour tout $t\in[t_0; t_1]$:
$$
\overline{S}_G(t)\geqslant \overline{S}_G(t_0) \exp\left(-\int_{t_0}^t F(s, I(s))ds \right) .
$$
Donc, pour tout $t\in[t_0; t_1]$, nous avons $\overline{S}_G(t) > 0$.  \\
D'apr\`es la cinqui\`eme \'equation du syst\`eme (\ref{mp3}), en posant $t_0=0$ et $t_1=r_0>0$, pour tout temps $t\in[t_0; t_1]$, nous avons:
$$
\begin{array}{rl}
\dfrac{d\overline{I}_G}{dt} &= m_1I_G\overline{S}_G-\gamma \overline{I}_G+a_9\overline{I}_G-a_{10}\overline{I}_M+a_{11}\overline{S}_GI_M+a_{12}\overline{S}_MI_G \medskip \\
               &\geqslant -\gamma \overline{I}_G+a_9\overline{I}_G-a_{10}\overline{I}_M \ \ \mbox{ (car, $m_1 > 0$ et, pour tout $t\in\mathbb{R}_+$, $a_{11} \geqslant 0$, $a_{12} \geqslant 0$)} \medskip \\
               & \geqslant 
\left\{              
\begin{array}{ll}
-\gamma \overline{I}_G-a_{10}\overline{I}_M & \mbox{ si } \ \varphi(t) >0  \medskip \\
-\gamma \overline{I}_G+a_9\overline{I}_G & \mbox{ si } \ \varphi(t) <0 \medskip \\
\end{array}
\right. \medskip \\
               &= 
\left\{              
\begin{array}{ll}
-(\gamma \overline{I}_G+a_{10}\overline{I}_M) & \mbox{ si } \ \varphi(t) >0  \medskip \\
-(\gamma -a_9)\overline{I}_G & \mbox{ si } \ \varphi(t) <0 \medskip \\
\end{array}
\right. \medskip \\ 
              &\geqslant
\left\{              
\begin{array}{ll}
-(\gamma +a_{10})\overline{I}_G & \mbox{ si } \ \varphi(t) >0  \medskip \\
-(\gamma -a_9)\overline{I}_G & \mbox{ si } \ \varphi(t) <0 \medskip \\
\end{array}
\right. \medskip \\ 
& \ \ \ (\mbox{en prenant arbitrairement $\overline{I}_G(t_0)>\overline{I}_M(t_0)$ et donc, pour tout $t\in[t_0; t_1]$, $\overline{I}_G(t)>\overline{I}_M(t)$}). \medskip \\
\end{array}
$$
D'o\`u, pour tout $t\in[t_0; t_1]$, nous avons:
$$\left\{              
\begin{array}{ll}
\dfrac{d\overline{I}_G}{dt}+(\gamma+a_{10})\overline{I}_G \geqslant 0 & \mbox{ si } \ \varphi(t) >0  \medskip \\
\dfrac{d\overline{I}_G}{dt}+(\gamma-a_9)\overline{I}_G \geqslant 0 & \mbox{ si } \ \varphi(t) <0 \medskip \\
\end{array}
\right. 
$$
Nous posons:
$$
F(t)=\min_{t\in[t_0; t_1]}
\left\{              
\begin{array}{ll}
\gamma+a_{10} & \mbox{ si } \ \varphi(t) >0  \medskip \\
\gamma-a_9 & \mbox{ si } \ \varphi(t) <0 \medskip \\
\end{array}
\right. \\ 
$$
Alors, pour tout $t\in[t_0; t_1]$, nous avons:
$$
\dfrac{d\overline{I}_G}{dt}+\overline{I}_G(t)F(t) \geqslant 0 .
$$
Pour tout $t\in[t_0; t_1]$, en multipliant les deux membres de l'in\'egalit\'e par $\exp \left(\displaystyle\int_{t_0}^t F(s)ds \right)>0$, nous obtenons:
$$
\exp\left(\int_{t_0}^t F(s)ds \right)\times\dfrac{d\overline{I}_G(t)}{dt} + F(t)\exp\left(\displaystyle\int_{t_0}^t F(s)ds \right)\times \overline{I}_G(t) \geqslant 0 .
$$
Soit, pour tout $t\in[t_0; t_1]$:
$$
\dfrac{d}{dt}\left[\overline{I}_G(t)\times \exp\left(\int_{t_0}^t F(s)ds \right)\right] \geqslant 0 .
$$
Ensuite, pour tout $t\in[t_0; t_1]$, par int\'egration entre $t_0$ et $t$, nous avons:
$$
\int_{t_0}^t \dfrac{d}{ds}\left[\overline{I}_G(s) \exp\left(\int_{t_0}^t F(s)ds \right)\right] ds \geqslant 0 .
$$
Nous en d\'eduisons que, pour tout $t\in[t_0; t_1]$:
$$
\overline{I}_G(t)\geqslant \overline{I}_G(t_0) \exp\left(-\int_{t_0}^t F(s)ds \right) .
$$
Donc, pour tout $t\in[t_0; t_1]$, nous avons $\overline{I}_G(t) > 0$.  \\
D'apr\`es la sixi\`eme \'equation du syst\`eme (\ref{mp3}), en posant $t_0=0$ et $t_1=r_0>0$, pour tout temps $t\in[t_0; t_1]$, nous avons:
$$
\begin{array}{rl}
\dfrac{d\overline{R}_G}{dt} &= \gamma \overline{I}_G+a_1\overline{R}_G-a_2\overline{R}_M \medskip \\
               &\geqslant a_1\overline{R}_G-a_2\overline{R}_M \ \ \mbox{ (car $\gamma > 0$)} \medskip \\
               & \geqslant 
\left\{              
\begin{array}{ll}
-a_2\overline{R}_M & \mbox{ si } \ \varphi(t) >0  \medskip \\
a_1\overline{R}_G & \mbox{ si } \ \varphi(t) <0 \medskip \\
\end{array}
\right. \medskip \\
               &\geqslant
\left\{              
\begin{array}{ll}
-a_2\overline{R}_G & \mbox{ si } \ \varphi(t) >0  \medskip \\
a_1\overline{R}_G & \mbox{ si } \ \varphi(t) <0 \medskip \\
\end{array}
\right. \medskip \\ 
& \ \ \ (\mbox{en supposant arbitrairement que $N_G>N_M$ et donc, pour tout $t\in[t_0; t_1]$, $\overline{R}_G(t)>\overline{R}_M(t)$}). \medskip \\
\end{array}
$$
D'o\`u, pour tout $t\in[t_0; t_1]$, nous avons:
$$\left\{              
\begin{array}{ll}
\dfrac{d\overline{R}_G}{dt}+a_2\overline{R}_G \geqslant 0 & \mbox{ si } \ \varphi(t) >0  \medskip \\
\dfrac{d\overline{R}_G}{dt}-a_1\overline{R}_G \geqslant 0 & \mbox{ si } \ \varphi(t) <0 \medskip \\
\end{array}
\right. 
$$
Nous posons:
$$
F(t)=\min_{t\in[t_0; t_1]}
\left\{              
\begin{array}{ll}
a_2 & \mbox{ si } \ \varphi(t) >0  \medskip \\
-a_1 & \mbox{ si } \ \varphi(t) <0 \medskip \\
\end{array}
\right. \\ 
$$
Alors, pour tout $t\in[t_0; t_1]$, nous avons:
$$
\dfrac{d\overline{R}_G}{dt}+\overline{R}_G(t)F(t) \geqslant 0 .
$$
Pour tout $t\in[t_0; t_1]$, en multipliant les deux membres de l'in\'egalit\'e par $\exp \left(\displaystyle\int_{t_0}^t F(s)ds \right)>0$, nous obtenons:
$$
\exp\left(\int_{t_0}^t F(s)ds \right)\times\dfrac{d\overline{R}_G(t)}{dt} + F(t)\exp\left(\displaystyle\int_{t_0}^t F(s)ds \right)\times \overline{R}_G(t) \geqslant 0 .
$$
Soit, pour tout $t\in[t_0; t_1]$:
$$
\dfrac{d}{dt}\left[\overline{R}_G(t)\times \exp\left(\int_{t_0}^t F(s)ds \right)\right] \geqslant 0 .
$$
Ensuite, pour tout $t\in[t_0; t_1]$, par int\'egration entre $t_0$ et $t$, nous avons:
$$
\int_{t_0}^t \dfrac{d}{ds}\left[\overline{R}_G(s) \exp\left(\int_{t_0}^t F(s)ds \right)\right] ds \geqslant 0 .
$$
Nous en d\'eduisons que, pour tout $t\in[t_0; t_1]$:
$$
\overline{R}_G(t)\geqslant \overline{R}_G(t_0) \exp\left(-\int_{t_0}^t F(s)ds \right) .
$$
Donc, pour tout $t\in[t_0; t_1]$, nous avons $\overline{R}_G(t) \geqslant 0$.  \\
Par sym\'etrie, nous d\'eduisons que, pour tout $t\in[t_0; t_1]$:
$$
S_M(t) > 0, \ I_M(t) > 0,\  R_M(t) \geqslant 0, \ \overline{S}_M(t) > 0, \ \overline{I}_M(t) > 0, \ \overline{R}_M(t) \geqslant 0 .
$$
Maintenant, nous supposons que, pour tout temps $t\in[t_0; t_1]$:
$$
\forall j\in \{G; M \}, \ S_j(t) > 0, \ I_j(t) > 0,\  R_j(t) \geqslant 0, \ \overline{S}_j(t) > 0, \ \overline{I}_j(t) > 0, \ \overline{R}_j(t) \geqslant 0 .
$$
Alors, il existe un certain instants $r_1\in\mathbb{R}_+^\ast$ tel que, pendant le temps $t\in[t_1; t_1+r_1]$, toutes les variables d'\'etat restent positives ou strictement positives jusqu'au moment o\`u l'une d'entre elles s'annule.  \\
Puisque les fonctions $\varphi$ et $\varphi_2$ sont $1-$p\'eriodiques, de la premi\`ere \'equation du syst\`eme (\ref{mp3}), par un raisonnement similaire au temps $t\in[t_0; t_1]$, en posant $t_2=t_1+r_1>0$, en prenant
$$
F(t, I(t))=\min_{t\in[t_1; t_2]}
\left\{              
\begin{array}{ll}
m_1I_G+a_3I_M+m_5\overline{I}_G+a_7\overline{I}_M +a_2+a_6 & \mbox{ si } \ \varphi(t) >0 \ \mbox{ et } \ \varphi_2(t)=2  \medskip \\
m_1I_G-a_1+a_3I_M+m_5\overline{I}_G-a_5+a_7\overline{I}_M & \mbox{ si } \ \varphi(t) <0 \ \mbox{ et } \ \varphi_2(t)=2 \medskip \\
m_1I_G+m_5\overline{I}_G+a_2+a_6 & \mbox{ si } \ \varphi(t) >0 \ \mbox{ et } \ \varphi_2(t) = 0  \medskip \\
m_1I_G-a_1+m_5\overline{I}_G-a_5 & \mbox{ si } \ \varphi(t) <0 \ \mbox{ et } \ \varphi_2(t) = 0 \medskip \\
\end{array}
\right. \\ 
$$
nous montons que, pour chaque instant $t\in[t_1; t_2]$,
$$
\dfrac{dS_G}{dt}+S_G(t)F(t, I(t)) \geqslant 0 .
$$
Puis, comme les fonctions $\varphi$ et $\varphi_2$ sont $1-$p\'eriodiques, en multipliant les deux membres de l'in\'egalit\'e par $\exp \left(\displaystyle\int_{t_1}^t F(s, I(s))ds \right)>0$, nous d\'eduisons de $S_G(t) > 0$ pour tout $t\in[t_0; t_1]$ que $S_G(t) > 0$ pour tout $t\in[t_1; t_2]$.  \\
Puisque les fonctions $\varphi$ et $\varphi_2$ sont $1-$p\'eriodiques, de la deuxi\`eme \'equation du syst\`eme (\ref{mp3}), par un raisonnement similaire au temps $t\in[t_0; t_1]$, en posant $t_2=t_1+r_1>0$, en prenant
$$
F(t)=\min_{t\in[t_1; t_2]}
\left\{              
\begin{array}{ll}
\gamma+a_{10} & \mbox{ si } \ \varphi(t) >0  \medskip \\
\gamma-a_9 & \mbox{ si } \ \varphi(t) <0 \medskip \\
\end{array}
\right. \\ 
$$
nous montons que, pour tout temps $t\in[t_1; t_2]$,
$$
\dfrac{dI_G}{dt}+I_G(t)F(t, I(t)) \geqslant 0 .
$$
Puis, comme les fonctions $\varphi$ et $\varphi_2$ sont $1-$p\'eriodiques, en multipliant les deux membres de l'in\'egalit\'e par $\exp \left(\displaystyle\int_{t_1}^t F(s)ds \right)>0$, nous d\'eduisons de $I_G(t) > 0$ pour tout $t\in[t_0; t_1]$ que $I_G(t) > 0$ pour tout $t\in[t_1; t_2]$.  \\
Puisque les fonctions $\varphi$ et $\varphi_2$ sont $1-$p\'eriodiques, de la troisi\`eme \'equation du syst\`eme (\ref{mp3}), par un raisonnement similaire au temps $t\in[t_0; t_1]$, en posant $t_2=t_1+r_1>0$, en prenant
$$
F(t)=\min_{t\in[t_1; t_2]}
\left\{              
\begin{array}{ll}
a_2 & \mbox{ si } \ \varphi(t) >0  \medskip \\
-a_1 & \mbox{ si } \ \varphi(t) <0 \medskip \\
\end{array}
\right. \\ 
$$
nous montons que, pour tout temps $t\in[t_1; t_2]$,
$$
\dfrac{dR_G}{dt}+R_G(t)F(t) \geqslant 0 .
$$
Puis, comme les fonctions $\varphi$ et $\varphi_2$ sont $1-$p\'eriodiques, en multipliant les deux membres de l'in\'egalit\'e par $\exp \left(\displaystyle\int_{t_1}^t F(s)ds \right)>0$, nous d\'eduisons de $R_G(t) \geqslant 0$ pour tout $t\in[t_0; t_1]$ que $R_G(t) \geqslant 0$ pour tout $t\in[t_1; t_2]$.  \\
Puisque les fonctions $\varphi$ et $\varphi_2$ sont $1-$p\'eriodiques, de la quatri\`eme \'equation du syst\`eme (\ref{mp3}), par un raisonnement similaire au temps $t\in[t_0; t_1]$, en posant $t_2=t_1+r_1>0$, en prenant
$$
F(t, I(t))=\min_{t\in[t_1; t_2]}
\left\{              
\begin{array}{ll}
m_1I_G+a_3I_M+m_5\overline{I}_G+a_7\overline{I}_M+a_2+a_6 & \mbox{ si } \ \varphi(t) >0 \ \mbox{ et } \ \varphi_2(t)=2  \medskip \\
m_1I_G-a_1+a_3I_M+m_5\overline{I}_G-a_5+a_7\overline{I}_M & \mbox{ si } \ \varphi(t) <0 \ \mbox{ et } \ \varphi_2(t)=2 \medskip \\
m_1I_G+m_5\overline{I}_G+a_2+a_6 & \mbox{ si } \ \varphi(t) >0 \ \mbox{ et } \ \varphi_2(t) = 0  \medskip \\
m_1I_G-a_1+m_5\overline{I}_G-a_5 & \mbox{ si } \ \varphi(t) <0 \ \mbox{ et } \ \varphi_2(t) = 0 \medskip \\
\end{array}
\right. \\ 
$$
nous montons que, pour tout temps $t\in[t_1; t_2]$,
$$
\dfrac{d\overline{S}_G}{dt}+\overline{S}_G(t)F(t, I(t)) \geqslant 0 .
$$
Puis, comme les fonctions $\varphi$ et $\varphi_2$ sont $1-$p\'eriodiques, en multipliant les deux membres de l'in\'egalit\'e par $\exp \left(\displaystyle\int_{t_1}^t F(s, I(s))ds \right)>0$, nous d\'eduisons de $\overline{S}_G(t) > 0$ pour tout $t\in[t_0; t_1]$ que $\overline{S}_G(t) > 0$ pour tout $t\in[t_1; t_2]$.  \\
Puisque les fonctions $\varphi$ et $\varphi_2$ sont $1-$p\'eriodiques, de la cinqui\`eme \'equation du syst\`eme (\ref{mp3}), par un raisonnement similaire au temps $t\in[t_0; t_1]$, en posant $t_2=t_1+r_1>0$, en prenant
$$
F(t)=\min_{t\in[t_1; t_2]}
\left\{              
\begin{array}{ll}
\gamma+a_{10} & \mbox{ si } \ \varphi(t) >0  \medskip \\
\gamma-a_9 & \mbox{ si } \ \varphi(t) <0 \medskip \\
\end{array}
\right. \\ 
$$
nous montons que, pour tout temps $t\in[t_1; t_2]$,
$$
\dfrac{d\overline{I}_G}{dt}+\overline{I}_G(t)F(t) \geqslant 0 .
$$
Puis, comme les fonctions $\varphi$ et $\varphi_2$ sont $1-$p\'eriodiques, en multipliant les deux membres de l'in\'egalit\'e par $\exp \left(\displaystyle\int_{t_1}^t F(s)ds \right)>0$, nous d\'eduisons de $\overline{I}_G(t) > 0$ pour tout $t\in[t_0; t_1]$ que $\overline{I}_G(t) > 0$ pour tout $t\in[t_1; t_2]$.  \\
Puisque les fonctions $\varphi$ et $\varphi_2$ sont $1-$p\'eriodiques, de la sixi\`eme \'equation du syst\`eme (\ref{mp3}), par un raisonnement similaire au temps $t\in[t_0; t_1]$, en posant $t_2=t_1+r_1>0$, en prenant
$$
F(t)=\min_{t\in[t_1; t_2]}
\left\{              
\begin{array}{ll}
a_2 & \mbox{ si } \ \varphi(t) >0  \medskip \\
-a_1 & \mbox{ si } \ \varphi(t) <0 \medskip \\
\end{array}
\right. \\ 
$$
nous montons que, pour tout temps $t\in[t_1; t_2]$,
$$
\dfrac{d\overline{R}_G}{dt}+\overline{R}_G(t)F(t) \geqslant 0 .
$$
Puis, comme les fonctions $\varphi$ et $\varphi_2$ sont $1-$p\'eriodiques, en multipliant les deux membres de l'in\'egalit\'e par $\exp \left(\displaystyle\int_{t_1}^t F(s)ds \right)>0$, nous d\'eduisons de $\overline{R}_G(t) \geqslant 0$ pour tout $t\in[t_0; t_1]$ que $\overline{R}_G(t) \geqslant 0$ pour tout $t\in[t_1; t_2]$.  \\
Par sym\'etrie, nous d\'eduisons que, pour tout $t\in[t_1; t_2]$:
$$
S_M(t) > 0, \ I_j(t) > 0,\  R_M(t) \geqslant 0, \ \overline{S}_M(t) > 0, \ \overline{I}_M(t) > 0, \ \overline{R}_M(t) \geqslant 0 .
$$
Les fonctions $\varphi$ et $\varphi_2$ \'etant p\'eriodiques de p\'eriode $1$. Par r\'ecurence, montrons que:  \\
si, pour tout $n\in\mathbb{N}\backslash \{0; 1 \}$ et pour tout temps $t\in[t_{n-2}; t_{n-1}]$,  
$$
\forall j\in \{G; M \}, \ S_j(t) > 0, \ I_j(t) > 0,\  R_j(t) \geqslant 0, \ \overline{S}_j(t) > 0, \ \overline{I}_j(t) > 0, \ \overline{R}_j(t) \geqslant 0 
$$
alors, pour tout temps $t\in[t_{n-1}; t_n]$, nous avons 
$$
\forall j\in \{G; M \}, \ S_j(t) > 0, \ I_j(t) > 0,\  R_j(t) \geqslant 0, \ \overline{S}_j(t) > 0, \ \overline{I}_j(t) > 0, \ \overline{R}_j(t) \geqslant 0 .
$$
Le r\'esultat est vrai pour $n=2$ car, d'apr\`es ce qui pr\'ec\`ede, si, pour tout temps $t\in[t_0; t_1]$, 
$$
\forall j\in \{G; M \}, \ S_j(t) > 0, \ I_j(t) > 0,\  R_j(t) \geqslant 0, \ \overline{S}_j(t) > 0, \ \overline{I}_j(t) > 0, \ \overline{R}_j(t) \geqslant 0 
$$
alors, pour tout temps $t\in[t_1; t_2]$, nous avons
$$
\forall j\in \{G; M \}, \ S_j(t) > 0, \ I_j(t) > 0,\  R_j(t) \geqslant 0, \ \overline{S}_j(t) > 0, \ \overline{I}_j(t) > 0, \ \overline{R}_j(t) \geqslant 0 .
$$
Soit $n\in\mathbb{N}\backslash \{0; 1 \}$. Supposons que le r\'esultat est vrai jusqu'au rang $n$. \\
Montrons le jusqu'au rang $n+1$. \\
Soit, pour $n\in\mathbb{N}^\ast$, un temps $t\in[t_{n-1}; t_n]$ tels que:
$$
\forall j\in \{G; M \}, \ S_j(t) > 0, \ I_j(t) > 0,\  R_j(t) \geqslant 0, \ \overline{S}_j(t) > 0, \ \overline{I}_j(t) > 0, \ \overline{R}_j(t) \geqslant 0 .
$$
Alors, pour $n\in\mathbb{N}$, il existe un un certain instants $r_n\in\mathbb{R}_+^\ast$ tel que, pendant le temps $t\in[t_n; t_n+r_n]$, tous les compartiments restent positifs ou strictement positifs jusqu'au moment o\`u l'un d'entre eux s'annule.  \\
Puisque les fonctions $\varphi$ et $\varphi_2$ sont $1-$p\'eriodiques, de la premi\`ere \'equation du syst\`eme (\ref{mp3}), par un raisonnement similaire au temps $t\in[t_0; t_1]$, pour $n\in\mathbb{N}$, en posant $t_{n+1}=t_n+r_n>0$, en prenant
$$
F(t, I(t))=\min_{t\in[t_n; t_{n+1}]}
\left\{              
\begin{array}{ll}
m_1I_G+a_3I_M+m_5\overline{I}_G+a_7\overline{I}_M +a_2+a_6 & \mbox{ si } \ \varphi(t) >0 \ \mbox{ et } \ \varphi_2(t)=2  \medskip \\
m_1I_G-a_1+a_3I_M+m_5\overline{I}_G-a_5+a_7\overline{I}_M & \mbox{ si } \ \varphi(t) <0 \ \mbox{ et } \ \varphi_2(t)=2 \medskip \\
m_1I_G+m_5\overline{I}_G+a_2+a_6 & \mbox{ si } \ \varphi(t) >0 \ \mbox{ et } \ \varphi_2(t) = 0  \medskip \\
m_1I_G-a_1+m_5\overline{I}_G-a_5 & \mbox{ si } \ \varphi(t) <0 \ \mbox{ et } \ \varphi_2(t) = 0 \medskip \\
\end{array}
\right. \\ 
$$
nous d\'eduisons que, pour chaque instant $t\in[t_n; t_{n+1}]$,
$$
\dfrac{dS_G}{dt}+S_G(t)F(t, I(t)) \geqslant 0 .
$$
Par suite, comme les fonctions $\varphi$ et $\varphi_2$ sont $1-$p\'eriodiques, pour $n\in\mathbb{N}$, en multipliant les deux membres de l'in\'egalit\'e par $\exp \left(\displaystyle\int_{t_n}^t F(s, I(s))ds \right)>0$, nous d\'eduisons de hypoth\`ese de r\'ecurrence que $S_G(t) > 0$ pour tout $t\in[t_n; t_{n+1}]$.  \\
Puisque les fonctions $\varphi$ et $\varphi_2$ sont $1-$p\'eriodiques, de la deuxi\`eme \'equation du syst\`eme (\ref{mp3}), par un raisonnement similaire au temps $t\in[t_0; t_1]$, pour $n\in\mathbb{N}$, en posant $t_{n+1}=t_n+r_n>0$, en prenant
$$
F(t)=\min_{t\in[t_n; t_{n+1}]}
\left\{              
\begin{array}{ll}
\gamma+a_{10} & \mbox{ si } \ \varphi(t) >0  \medskip \\
\gamma-a_9 & \mbox{ si } \ \varphi(t) <0 \medskip \\
\end{array}
\right. \\ 
$$
nous d\'eduisons que, pour tout temps $t\in[t_n; t_{n+1}]$,
$$
\dfrac{dI_G}{dt}+I_G(t)F(t) \geqslant 0 .
$$
Par suite, comme les fonctions $\varphi$ et $\varphi_2$ sont $1-$p\'eriodiques, pour $n\in\mathbb{N}$, en multipliant les deux membres de l'in\'egalit\'e par $\exp \left(\displaystyle\int_{t_n}^t F(s)ds \right)>0$, nous d\'eduisons de hypoth\`ese de r\'ecurrence que $I_G(t) > 0$ pour tout $t\in[t_n; t_{n+1}]$.  \\
Puisque les fonctions $\varphi$ et $\varphi_2$ sont $1-$p\'eriodiques, de la troisi\`eme \'equation du syst\`eme (\ref{mp3}), par un raisonnement similaire au temps $t\in[t_0; t_1]$, pour $n\in\mathbb{N}$, en posant $t_{n+1}=t_n+r_n>0$, en prenant
$$
F(t)=\min_{t\in[t_n; t_{n+1}]}
\left\{              
\begin{array}{ll}
a_2 & \mbox{ si } \ \varphi(t) >0  \medskip \\
-a_1 & \mbox{ si } \ \varphi(t) <0 \medskip \\
\end{array}
\right. \\ 
$$
nous d\'eduisons que, pour tout temps $t\in[t_n; t_{n+1}]$,
$$
\dfrac{dR_G}{dt}+R_G(t)F(t) \geqslant 0 .
$$
Par suite, comme les fonctions $\varphi$ et $\varphi_2$ sont $1-$p\'eriodiques, pour $n\in\mathbb{N}$, en multipliant les deux membres de l'in\'egalit\'e par $\exp \left(\displaystyle\int_{t_n}^t F(s)ds \right)>0$, nous d\'eduisons de hypoth\`ese de r\'ecurrence que $R_G(t) > 0$ pour tout $t\in[t_n; t_{n+1}]$.  \\
Puisque les fonctions $\varphi$ et $\varphi_2$ sont $1-$p\'eriodiques, de la quatri\`eme \'equation du syst\`eme (\ref{mp3}), par un raisonnement similaire au temps $t\in[t_0; t_1]$, pour $n\in\mathbb{N}$, en posant $t_{n+1}=t_n+r_n>0$, en prenant
$$
F(t, I(t))=\min_{t\in[t_n; t_{n+1}]}
\left\{              
\begin{array}{ll}
m_1I_G+a_3I_M+m_5\overline{I}_G+a_7\overline{I}_M+a_2+a_6 & \mbox{ si } \ \varphi(t) >0 \ \mbox{ et } \ \varphi_2(t)=2  \medskip \\
m_1I_G-a_1+a_3I_M+m_5\overline{I}_G-a_5+a_7\overline{I}_M & \mbox{ si } \ \varphi(t) <0 \ \mbox{ et } \ \varphi_2(t)=2 \medskip \\
m_1I_G+m_5\overline{I}_G+a_2+a_6 & \mbox{ si } \ \varphi(t) >0 \ \mbox{ et } \ \varphi_2(t) = 0  \medskip \\
m_1I_G-a_1+m_5\overline{I}_G-a_5 & \mbox{ si } \ \varphi(t) <0 \ \mbox{ et } \ \varphi_2(t) = 0 \medskip \\
\end{array}
\right. \\ 
$$
nous d\'eduisons que, pour tout temps $t\in[t_n; t_{n+1}]$,
$$
\dfrac{d\overline{S}_G}{dt}+\overline{S}_G(t)F(t) \geqslant 0 .
$$
Par suite, comme les fonctions $\varphi$ et $\varphi_2$ sont $1-$p\'eriodiques, pour $n\in\mathbb{N}$, en multipliant les deux membres de l'in\'egalit\'e par $\exp \left(\displaystyle\int_{t_n}^t F(s, I(s))ds \right)>0$, nous d\'eduisons de hypoth\`ese de r\'ecurrence que $\overline{S}_G(t) > 0$ pour tout $t\in[t_n; t_{n+1}]$.  \\
Puisque les fonctions $\varphi$ et $\varphi_2$ sont $1-$p\'eriodiques, de la cinqui\`eme \'equation du syst\`eme (\ref{mp3}), par un raisonnement similaire au temps $t\in[t_0; t_1]$, pour $n\in\mathbb{N}$, en posant $t_{n+1}=t_n+r_n>0$, en prenant
$$
F(t)=\min_{t\in[t_n; t_{n+1}]}
\left\{              
\begin{array}{ll}
\gamma+a_{10} & \mbox{ si } \ \varphi(t) >0  \medskip \\
\gamma-a_9 & \mbox{ si } \ \varphi(t) <0 \medskip \\
\end{array}
\right. \\ 
$$
nous d\'eduisons que, pour tout temps $t\in[t_n; t_{n+1}]$,
$$
\dfrac{d\overline{I}_G}{dt}+\overline{I}_G(t)F(t) \geqslant 0 .
$$
Par suite, comme les fonctions $\varphi$ et $\varphi_2$ sont $1-$p\'eriodiques, pour $n\in\mathbb{N}$, en multipliant les deux membres de l'in\'egalit\'e par $\exp \left(\displaystyle\int_{t_n}^t F(s)ds \right)>0$, nous d\'eduisons de hypoth\`ese de r\'ecurrence que $\overline{I}_G(t) > 0$ pour tout $t\in[t_n; t_{n+1}]$.  \\
Puisque les fonctions $\varphi$ et $\varphi_2$ sont $1-$p\'eriodiques, de la sixi\`eme \'equation du syst\`eme (\ref{mp3}), par un raisonnement similaire au temps $t\in[t_0; t_1]$, pour $n\in\mathbb{N}$, en posant $t_{n+1}=t_n+r_n>0$, en prenant
$$
F(t)=\min_{t\in[t_n; t_{n+1}]}
\left\{              
\begin{array}{ll}
a_2 & \mbox{ si } \ \varphi(t) >0  \medskip \\
-a_1 & \mbox{ si } \ \varphi(t) <0 \medskip \\
\end{array}
\right. \\ 
$$
nous d\'eduisons que, pour tout temps $t\in[t_n; t_{n+1}]$,
$$
\dfrac{d\overline{R}_G}{dt}+\overline{R}_G(t)F(t) \geqslant 0 .
$$
Par suite, comme les fonctions $\varphi$ et $\varphi_2$ sont $1-$p\'eriodiques, pour $n\in\mathbb{N}$, en multipliant les deux membres de l'in\'egalit\'e par $\exp \left(\displaystyle\int_{t_n}^t F(s)ds \right)>0$, nous d\'eduisons de hypoth\`ese de r\'ecurrence que $\overline{R}_G(t) \geqslant 0$ pour tout $t\in[t_n; t_{n+1}]$.  \\
Par sym\'etrie, nous d\'eduisons que, si, pour $n\in\mathbb{N}^\ast$ et pour tout temps $t\in[t_{n-1}; t_n]$,
$$
S_M(t) > 0,\ I_M(t) > 0,\  R_M(t) \geqslant 0, \ \overline{S}_M(t) > 0, \ \overline{I}_M(t) > 0, \ \overline{R}_M(t) \geqslant 0 
$$
alors, pour tout temps $t\in[t_n; t_{n+1}]$, nous avons
$$
S_M(t) > 0,\ I_M(t) > 0,\  R_M(t) \geqslant 0, \ \overline{S}_M(t) > 0, \ \overline{I}_M(t) > 0, \ \overline{R}_M(t) \geqslant 0 . 
$$
Il en r\'esulte que le r\'esultat est vrai au rang $n+1$. Ce qui ach\`eve la r\'ecurrence.  \\
Finalement, d\`es que toutes les conditions initiales sont positives ou strictement positives, nous avons bien:
$$
\forall j\in \{G; M \}, \ \forall t \in \mathbb{R}_+, \ S_j \geqslant 0, \ I_j \geqslant 0,\  R_j \geqslant 0, \ \overline{S}_j \geqslant 0, \ \overline{I}_j \geqslant 0, \ \overline{R}_j \geqslant 0 .
$$
Montrons que $\Omega$ est un ensemble positivement invariant. \\
En additionnant les six premi\`eres \'equations et les six derni\`eres \'equations du syst\`eme (\ref{mp3}), la population totale de chaque r\'egion varie selon les \'equations diff\'erentielles ordinaires: 
$$
\begin{array}{rl}
\dfrac{dN_G}{dt}&=(a_1+a_5)S_G+a_9I_G+a_1R_G+(a_1+a_5)\overline{S}_G+a_9\overline{I}_G+a_1\overline{R}_G  \medskip \\
                &-(a_2+a_6)S_M-a_{10}I_M-a_2R_M-(a_2+a_6)\overline{S}_M-a_{10}\overline{I}_M-a_2\overline{R}_M  \medskip \\
                &+(a_{11}-a_3)S_GI_M+(a_{11}-a_3)\overline{S}_GI_M+(a_4+a_{12})S_MI_G+(a_4+a_{12})\overline{S}_MI_G  \medskip \\
                &-m_5S_G\overline{I}_G-m_5\overline{S}_G\overline{I}_G-a_7S_G\overline{I}_M-a_7\overline{S}_G\overline{I}_M+a_8S_M\overline{I}_G+a_8\overline{S}_M\overline{I}_G  \\
\end{array}
$$
et 
$$
\begin{array}{rl}
\dfrac{dN_M}{dt}&=(a_2+a_6)S_M+a_{10}I_M+a_2R_M+(a_2+a_6)\overline{S}_M+a_{10}\overline{I}_M+a_2\overline{R}_M  \medskip \\
                &-(a_1+a_5)S_G-a_9I_G-a_1R_G-(a_1+a_5)\overline{S}_G-a_9\overline{I}_G-a_1\overline{R}_G  \medskip \\
                &+(a_3+a_{11})S_GI_M+(a_3+a_{11})\overline{S}_GI_M+(a_{12}-a_4)S_MI_G+(a_{12}-a_4)\overline{S}_MI_G \medskip \\
                &-m_6S_M\overline{I}_M-m_6\overline{S}_M\overline{I}_M+a_3S_G\overline{I}_M+a_3\overline{S}_G\overline{I}_M-a_4S_M\overline{I}_G-a_4\overline{S}_M\overline{I}_G . \medskip \\
\end{array}
$$
Posons $a=\displaystyle\min_{t\in\mathbb{R}_+} \{a_2+a_6; a_{10}; a_2 \}<0$ et $a'=\displaystyle\min_{t\in\mathbb{R}_+} \{a_1+a_5; a_9; a_1 \}<0$.  \\
Alors, d\`es que les conditions
$$
\begin{array}{cc}
(a_{11}-a_3)(S_GI_M+\overline{S}_GI_M)+(a_4+a_{12})(S_MI_G+\overline{S}_MI_G)   \\
\leqslant \\
m_5(S_G\overline{I}_G+\overline{S}_G\overline{I}_G)+a_7(S_G\overline{I}_M+\overline{S}_G\overline{I}_M)-a_8(S_M\overline{I}_G+a_8\overline{S}_M\overline{I}_G) \\
\end{array}
$$
et 
$$
\begin{array}{cc}
(a_3+a_{11})(S_GI_M+\overline{S}_GI_M)+(a_{12}-a_4)(S_MI_G+\overline{S}_MI_G)  \\
\leqslant  \\
m_6(S_M\overline{I}_M+\overline{S}_M\overline{I}_M)-a_3(S_G\overline{I}_M+\overline{S}_G\overline{I}_M)+a_4(S_M\overline{I}_G+\overline{S}_M\overline{I}_G)  \\
\end{array}
$$
sont v\'erifi\'ees, nous obtenons: 
$$
\dfrac{dN_G}{dt}\leqslant aN_G \ \mbox{ et } \ \dfrac{dN_M}{dt} \leqslant a' N_M .
$$
Consid\'erons les \'equations diff\'erentielles ordinaires:
$$
\left\{              
\begin{array}{ll}
\widetilde{N}_G'(t) = a\widetilde{N}_G(t) \medskip \\
\widetilde{N}_G(0)=N_G(0) \medskip \\
\end{array}
\right. 
\ \mbox{ et } \ \
\left\{              
\begin{array}{ll}
\widetilde{N}_M'(t) = a\widetilde{N}_M(t) \medskip \\
\widetilde{N}_M(0)=N_M(0) \medskip \\
\end{array}
\right. 
$$
Alors, pour tout $t\in\mathbb{R}_+$, nous en d\'eduisons que:
$$
0\leqslant N_G(t) = N_G(0)e^{at} \leqslant N_G(0) \ \mbox{ et } \ 0\leqslant N_M(t) = N_M(0)e^{a't} \leqslant N_M(0) .
$$
Ce qui prouve la bornitude de $\Omega$. \\
On en d\'eduit que toutes les solutions des \'equations diff\'erentielles $N_G'(t) = aN_G(t)$ et $N_M'(t) =a' N_M(t)$ sont rest\'ees dans $\Omega$. \\
Donc, $\Omega$ est un ensemble positivement invariant.  \\
Montrons que $\Omega$ est un ferm\'e de $\mathbb{R}_+^{12}$.  \\
Par soucis de simplicit\'e, nous consid\'erons, pour $j\in\{G; M \}$, les proportions $s_j$, $i_j$, $r_j$, $\overline{s}_j$, $\overline{i}_j$ et $\overline{r}_j$. \\
Nous avons ainsi, pour $j\in\{G; M \}$:
$$
0 \leqslant s_j+i_j+r_j+\overline{s}_j+\overline{i}_j+\overline{r}_j \leqslant 1 .
$$
Puisque, pour $j\in \{G; M \}$, $s_j \geqslant 0$, $i_j \geqslant 0$,  $r_j \geqslant 0$, $\overline{s}_j \geqslant 0$, $\overline{i}_j \geqslant 0$, $\overline{r}_j \geqslant 0$, alors nous obtenons, pour $j\in \{G; M \}$, $(s_j, i_j, r_j, \overline{s}_j, \overline{i}_j, \overline{r}_j)\in [0; 1]^6$.  \\
Donc, $\Omega$ est un ferm\'e de $\mathbb{R}_+^{12}$ en tant que l'image r\'eciproque du ferm\'e $[0; 1]$ par l'application continue $$
(s_j, i_j, r_j, \overline{s}_j, \overline{i}_j, \overline{r}_j) \mapsto s_j+i_j+r_j+\overline{s}_j+\overline{i}_j+\overline{r}_j .
$$
Par suite, $\Omega$ est un compact de $\mathbb{R}_+^{12}$ comme ferm\'e et born\'e de $\mathbb{R}_+^{12}$.  \\
Finalement, nous en d\'eduisons que l'ensemble $\Omega$ est un compact positivement invariant de $\mathbb{R}_+^{12}$. 
\hfill $\Box$
\end{proof}

\section{Choix des param\`etres du mod\`ele}
\noindent Nous pr\'ecisons que nous n'avons pas l'intention de pr\'edire exactement les chiffres de l'\'epid\'emie mais d'analyser les effets des diff\'erentes parties du mod\`ele et dans quelle mesure ils peuvent repr\'esenter qualitativement les tendances de l'\'epid\'emie. Cela est d\^u au fait que les donn\'ees utilis\'ees ne repr\'esentent pas avec pr\'ecision l'incidence r\'eelle de l'\'epid\'emie, en raison de nombreux facteurs (cas non d\'etect\'es ou non signal\'es ou encore non pris en compte, $\cdots$). Ainsi, nos r\'esultats num\'eriques peuvent s'av\'erer parfois \'eloign\'ee de la r\'ealit\'e \`a cause de donn\'ees exp\'erimentales incompl\`etes ou erron\'ees pour la p\'eriode du 13 mai 2020 au 31 juillet 2022.  \\

\noindent Pour la r\'esolution num\'erique des mod\`eles (\ref{d}) et (\ref{mp2}), nous avons utilis\'e la fonction \og odeint \fg \ de la librairie \og scipy.integrate \fg \ de Python ou la fonction \og ode \fg \ du package \og deSolve \fg \ de R. Car les deux logiciels donnent le m\^eme r\'esultat lors des simulations num\'eriques.  \\

\newpage

\noindent Nous r\'esumons le choix des param\`etres du mod\`ele propos\'e dans le tableau suivant: 
\begin{table}[hbtp]
\begin{center}
\begin{tabular}{|c|p{5cm}|p{4.75cm}|c|}
\hline
{\bf Param\`etres} & {\bf Description} & {\bf Valeurs} & {\bf R\'ef\'erences}  \\ 
\hline
$\gamma$ & Taux de gu\'erison & $\dfrac{1}{7}$ ou $\dfrac{1}{14}$ & \cite{Lauer and al. (2020), Wolfel and al. (2020)} \\
\hline
$\beta_G$ & Taux de transmission de Covid-19 en Guadeloupe & $0.14593$ ou $0.07297$ & Valeur estim\'ee  \\
\hline
$\beta_M$ & Taux de transmission de Covid-19 en Martinique & $0.14679$ ou $0.073398$ & Valeur estim\'ee  \\  
\hline
$\eta$ & Fraction des infect\'es se d\'epla\c{c}ant entre ces deux r\'egions & $[0,5; 0,9]$ & Hypoth\`ese  \\
\hline
$\tau$ & Nombre de journ\'ees pass\'ees dans chacune des r\'egions & $0,5$ & Hypoth\`ese  \\
\hline
$\varepsilon_i$ & Taux d'efficacit\'e int\'erieur (protection contre la maladie) & $\varepsilon_i (\mbox{tissu}) \in [0,20; 0,80]$ et $\varepsilon_i (\mbox{chirurgicaux}) \in [0,70; 0,90]$ & \cite{Eikenberry and al. (2020)} \\
\hline
$\varepsilon_e$ & Taux d'efficacit\'e ext\'erieur (protection contre la transmission de la maladie) & $\varepsilon_e (\mbox{tissu}) \in [0; 0,80]$ et $\varepsilon_e (\mbox{chirurgicaux}) \in [0,50; 0,90]$ & \cite{Eikenberry and al. (2020)} \\
\hline
$n_{GM}$ & Nombre d'individus se d\'epla\c{c}ant de la Guadeloupe vers la Martinique & $100$ & Hypoth\`ese  \\
\hline
$n_{MG}$ & Nombre d'individus se d\'epla\c{c}ant de la Martinique vers la Guadeloupe & $100$ & Hypoth\`ese  \\
\hline
NG & Population totale de la Guadeloupe & $376 \ 879$ & SPF \cite{SPF (2020)} \\
\hline
NM & Population totale de la Martinique & $358 \ 749$ & SPF \cite{SPF (2020)} \\
\hline
$\overline{S}_j$, $j\in \{G; M \}$ & Couverture de masques (fraction de la population initialement masqu\'ee) & Entre $70$\% de $N_j$ \`a $80$\% \ de $N_j$, avec $j\in \{G; M \}$ & Hypoth\`ese  \\
\hline
\end{tabular}
\caption{D\'efinition des param\`etres et leurs valeurs utilis\'ees pour les simulations num\'eriques du mod\`ele propos\'e.} \label{tdp}
\end{center} 
\end{table}

\section{Simulations num\'eriques du mod\`ele propos\'e}

\subsection{Effet de l'utilisation des masques sur la propagation de l'\'epid\'emie de Covid-19}\label{edum}
\noindent Nous supposons que les individus infect\'es ne portent pas de masques et que la propagation de l'\'epid\'emie est faite dans des conditions constantes (mobilit\'e sans restriction, pas de confinement, pas de distanciation sociale, pas de gestes barri\`eres, $\cdots$, hygi\`ene personnelle identique \`a celle pratiqu\'ee avant l'apparition de l'\'epid\'emie) sur l'intervalle de temps consid\'er\'e ($400$ jours). Les r\'esultats d'int\'er\^et sont le pic \'epid\'emique et le nombre cumul\'e de cas. Pour calibrer les conditions initiales du mod\`ele (\ref{mp2}), nous avons utilis\'e des valeurs ajust\'ees \`a partir des donn\'ees de l'\'epid\'emie de Covid-19 dans les r\'egions Guadeloupe et Martinique publi\'ees sur le site de  Sant\'e Publique France \cite{SPF (2020)}. Pour les simulations num\'eriques, nous supposons que $\varepsilon_i=\varepsilon_e=\varepsilon$. La valeur des autres param\`etres \'etant indiqu\'ees dans le tableau \ref{tdp} ci-dessus.  \\

\noindent Sur tous les graphiques ci-dessous, $SG$, $IG$ et $RG$ correspondent aux compartiments des individus non masqu\'es de la r\'egion de Guadeloupe; et $SGb$, $IGb$ et $RGb$ repr\'esentent ceux qui sont dans la classe masqu\'ee de cette m\^eme r\'egion. \\

\newpage

\begin{figure}[hbtp]
\centerline{\includegraphics[scale=1]{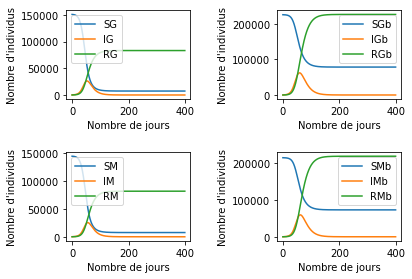}}
\caption{L'\'evolution du nombre d'individus dans chaque compartiment au bout de 400 jours lorsque $70$\% \ de la population portent des masques ayant un taux d'efficacit\'e globale de filtrage $\varepsilon=0.65$.}\label{f9}
\end{figure}

\noindent Nous supposons que $70$\% \ de la population de chaque r\'egion utilisent des masques et nous obtenons le graphique \ref{f9} \ ci-dessous.
 \\

\begin{figure}[hbtp]
\centerline{\includegraphics[scale=1]{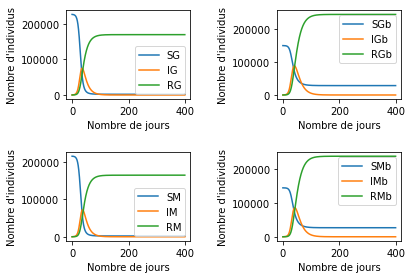}}
\caption{L'\'evolution du nombre d'individus dans chaque compartiment au bout de 400 jours lorsque $30$\% \ de la population utilisent des masques ayant un taux d'efficacit\'e globale $\varepsilon=0.65$.}\label{f18}
\end{figure}

\noindent Nous supposons maintenant que $30$\% \ de la population de chaque r\'egion portant de masques. Nous avons alors le graphique \ref{f18} \ suivant. 
 \\
Par lecture graphique, pour les deux niveaux de couverture, les individus susceptibles de la classe des non masqu\'ees semblent tous avoir \'et\'e touch\'es par l'\'epid\'emie de Covid-19 et une partie de ceux portant le masque semble \'epargner. Par ailleurs, suivant le niveau de couverture de masques, l'effet des masques sur la propagation de l'\'epid\'emie semble bien r\'eel. \\

\newpage

\begin{figure}[hbtp]
\centerline{\includegraphics[scale=1]{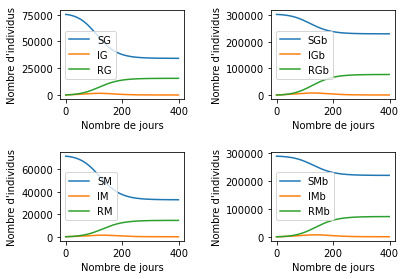}}
\caption{L'\'evolution du nombre d'individus dans chaque compartiment au bout de 400 jours lorsque $80$\% \ de la population utilisent des masques ayant un taux d'efficacit\'e globale $\varepsilon=0.65$.}\label{f19}
\end{figure}

\begin{figure}[hbtp]
\centerline{\includegraphics[scale=1]{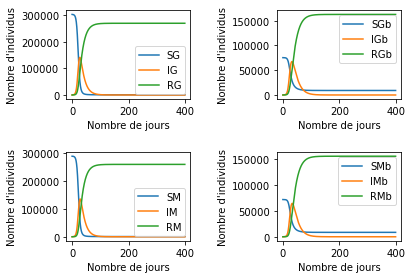}}
\caption{L'\'evolution du nombre d'individus dans chaque compartiment au bout de 400 jours lorsque $20$\% \ de la population utilisent des masques ayant un taux d'efficacit\'e globale $\varepsilon=0.65$.}\label{f20}
\end{figure}

\noindent Pour rendre cet effet plus visible, nous regardons deux autres niveaux de couverture des masques: $80$\% \ et $20$\%. \\
Si $80$\% \ la propagation utilisent le port du masque, alors nous avons le graphique \ref{f19} \ ci-dessous.
 \\
Si au contraire seulement $20$\% \ des sujets sont masqu\'es, alors nous obtenons le graphique \ref{f20} \ ci-dessous.
 \\
Les deux graphiques \ref{f19} \ et \ref{f20} \ semblent indiquer un effet tr\`es significatif pour l'utilisation des masques. En effet, le graphique \ref{f19} \ semble montrer qu'un grand nombre d'individus susceptibles (masqu\'es ou non masqu\'es) n'ont pas \'et\'e touch\'es par la Covid-19. De plus, les susceptibles de la classe des masqu\'es semblent moins infect\'es par la maladie. Il semble donc que la situation se stabilise (est sous contr\^ole), l'\'epid\'emie ne se propage pas et qu'elle finisse par s'\'eteindre au bout de quelques jours. Cela correspond \`a la stabilit\'e asymptotique de l'\'equilibre du syst\`eme. Par ailleurs, le graphique de la figure \ref{f20} \ semble d\'ecrire une autre situation de l'\'epid\'emie: tous les susceptibles semblent toucher au bout d'une centaine de jours par le virus avec une plus forte tendance pour les non masqu\'es, un pic \'epid\'emique atteint au bout d'une cinquantaine de jours mais plus important toutefois dans la classe des non masqu\'es, un nombre cumul\'e de cas plus important pour une couverture de $20$\%.

\subsection{Effet de la couverture et de l'efficacit\'e des masques sur la propagation de l'\'epid\'emie}
\noindent Nous examinons l'efficacit\'e et la couverture des masques (la fraction de la population portant habituellement de masques) comme deux param\`etres d'int\'er\^et pour un taux de contact infectieux fixe $\beta_G=\beta_M=0,5$. \\

\begin{figure}[hbtp]
\centerline{\includegraphics[scale=1]{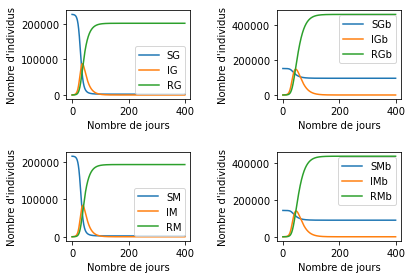}}
\caption{L'\'evolution du nombre d'individus dans chaque compartiment au bout de 400 jours lorsque $30$\% \ de la population utilisent des masques avec un taux d'efficacit\'e $\varepsilon=0.9$.}\label{f21}
\end{figure}

\noindent Nous supposons que $30$\% \ des individus de la population de chaque r\'egion portent le masque. Nous avons ainsi la figure \ref{f21} \ suivant. 
 \\
Comparer au r\'esultats obtenir avec la figure \ref{f18} \ lorsque nous avions $30$\% \ de couverture et un taux d'efficacit\'e $\varepsilon=0.65$, il semble que l'effet est moins visible lorsque nous augmentons le taux d'efficacit\'e des masques jusqu'\`a $\varepsilon=0.9$. \\

\begin{figure}[hbtp]
\centerline{\includegraphics[scale=1]{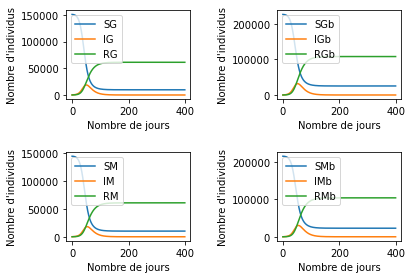}}
\caption{L'\'evolution du nombre d'individus dans chaque compartiment au bout de 400 jours lorsque $70$\% \ de la population utilisent des masques avec un taux d'efficacit\'e $\varepsilon=0,2$.}\label{f22}
\end{figure}

\noindent Nous consid\'erons une couverture de $70$\% \ et un taux d'efficacit\'e $\varepsilon=0,20$ pour la figure \ref{f22} \ ci-dessous.
 \\
L'effet semble beaucoup plus visible avec une meilleure couverture: comparer \`a la figure \ref{f9}, les susceptibles de la classe masqu\'ee semblent moins toucher par la maladie lorsqu'ils portaient un masque ayant un taux d'eficacit\'e $\varepsilon=0,65$. Par ailleurs, la situations des non masqu\'es semble ne pas changer. \\

\newpage

\noindent Nous regardons \`a pr\'esent une couverture de masques de $80$\% \ et un taux d'efficacit\'e $\varepsilon=0,20$.
\begin{figure}[hbtp]
\centerline{\includegraphics[scale=1]{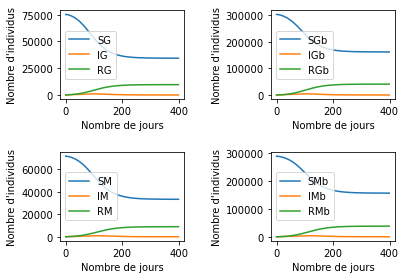}}
\caption{L'\'evolution du nombre d'individus dans chaque compartiment au bout de 400 jours lorsque $80$\% \ de la population utilisent des masques avec un taux d'efficacit\'e $\varepsilon=0,20$.}\label{f16}
\end{figure} \\
Les graphiques des figures \ref{f16} \ et \ref{f19} \ semblent montrer des effets tr\`es peu diff\'erents pour un m\^eme niveau de couverture avec des taux d'efficacit\'e $\varepsilon=0,20$ et $\varepsilon=0,65$. Il semble donc exister une l\'eg\`ere asym\'etrie entre la couverture et le taux d'efficacit\'e des masques: augmenter l\'eg\`erement la couverture de masques m\^eme peu efficaces semble en g\'en\'eral plus utile que d'augmenter le taux d'efficacit\'e des masques. Nous notons \'egalement, d'apr\`es la figure \ref{f16}, avec une couverture de masques de $80$\% \ et un taux d'efficacit\'e $\varepsilon \geqslant 0,20$, il semble que nous obtenons tr\`es peu de cas de Covid-19.  \\

\noindent Des mod\`eles math\'ematiques ant\'erieurs, li\'es par la grippe end\'emique, ont examin\'e l'utilit\'e du port du masque par le grand public \cite{Brienen and al. (2010), Tracht and al. (2010)}. Dans \cite{Tracht and al. (2010)}, Tracht et al. ont conclu que, pour la pand\'emie de la grippe H1N1, des masques peu efficaces (\`a $20$ \%) pourraient r\'eduire de moiti\'e le nombre d'infect\'es; tandis que si les masques n'\'etaient efficaces \`a int\'erieur qu'\`a $50$ \%, l'\'epid\'emie pourrait \^etre \'elimin\'ee si seulement $25$ \% \ de la population portaient des masques.  \\

\noindent Il est possible de repr\'esenter ces mesures sous forme de fonctions bidimensionnelles de couverture et d'efficacit\'e des masques en utilisant la matrice de corr\'elation de la couverture et de l'efficacit\'e de masques sous forme de cartes thermiques \cite{Eikenberry and al. (2020)}. \\

\noindent Nous ne savons pas dans quelle mesure les masques faits maison, g\'en\'eralement fabriqu\'es \`a partir de coton ou de torchon ou d'autres fibres de polyester, peuvent prot\'eger contre les gouttelettes ou a\'erosols et la transmission virale. Mais des r\'esultats exp\'erimentaux de Davies et al. \cite{Davies and al. (2013)} sugg\`erent que, m\^eme si les masques faits maison \'etaient moins efficaces que les masques chirurgicaux, ils \'etaient toujours nettement sup\'erieurs \`a l'absence de masque. Un essai clinique chez des professionnels de la sant\'e \cite{MacIntyre and al. (2015)} a montr\'e des performances relativement inf\'erieures pour les masques en tissu par rapport aux masques m\'edicaux.

\subsection{Effet de la couverture des masques et de la mobilit\'e sur la propagation de l'\'epid\'emie}
\noindent Nous simulons num\'eriquement une \'epid\'emie pour deux niveaux de couverture de masques et une mobilit\'e sans restriction sans modifier le taux de contact infectieux $\beta=0,5$ et le taux d'efficacit\'e global $\varepsilon=0,65$. Nous supposons la Guadeloupe compte $180$ individus infect\'es et que la Martinique en a $127$. \\

\newpage

\noindent S'il y a une couverture de masques de $70$\% \ dans chaque r\'egion et une mobilit\'e sans restriction entre elles, alors nous obtenons les graphiques suivants.  
\begin{figure}[hbtp]
\centerline{\includegraphics[scale=1]{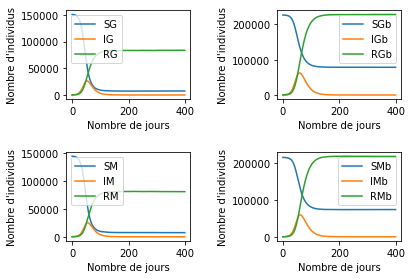}}
\caption{L'\'evolution des individus dans chaque compartiment au bout de 400 jours lorsque $70$\% de la population sont masqu\'ees avec une mobilit\'e sans restriction.}\label{f17}
\end{figure}  \\
S'il y a une couverture de masques de $30$\% \ dans chaque r\'egion et une mobilit\'e sans restriction entre elles, nous obtenons alors les graphiques suivants.  
\begin{figure}[hbtp]
\centerline{\includegraphics[scale=1]{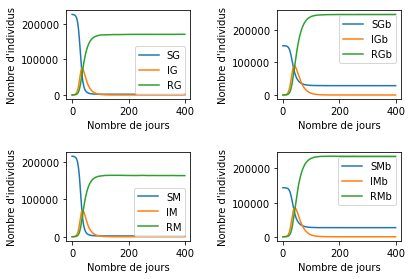}}
\caption{L'\'evolution des individus dans chaque compartiment au bout de 400 jours lorsque $30$\% de la population sont masqu\'ees avec une mobilit\'e sans restriction.}\label{f23}
\end{figure} \\
Nous pouvons remarquer la situation d\'ecrite par les figures \ref{f17} \ et \ref{f23} \ semble bien diff\'erente que celle o\`u il y a une mobilit\'e sans restriction: m\^eme avec une couverture de masques de $30$\%, la situation semble rester sous contr\^ole. Les figures \ref{f17} \ et \ref{f23} semblent donc indiquer que l'effet de la mobilit\'e sur la propagation de l'\'epid\'emie de Covid-19 est moins visible lorsque le port du masque est appliqu\'e par une fraction non n\'eglieable de la population.  \\

\noindent La relation entre la couverture et l'efficacit\'e des masques, le taux de contact infectieux et la mobilit\'e semblent non lin\'eaires. Cependant, des r\'eductions progressives du taux de contact infectieux $\beta$, en raison des mesures de distanciation sociale, $\cdots$, peuvent \^etre en synergie avec d'autres mesures de r\'eduction pour produire un effet significatif sur l'\'epid\'emie.

\section{Discussion et conclusion}
\noindent Les r\'esultats de nos simulations num\'eriques sugg\`erent que l'utilisation des masques faciaux dans la population peut retarder l'\'epid\'emie de Covid-19 et r\'eduire la transmission pour contenir l'\'epid\'emie. L'effet sur la taille finale de l'\'epid\'emie d\'epend de l'efficacit\'e du masque et de la couverture dans la population. \\

\noindent Des effets suppl\'ementaires, non inclus dans notre mod\`ele, pourraient rendre l'effet de l'utilisation de masques dans la population encore plus fort, comme par exemples:
\begin{enumerate}
\item[$\bullet$] l'utilisation de masques prot\`ege non seulement les individus en bonne sant\'e mais r\'eduit \'egalement la contagiosit\'e des porteurs symptomatiques et asymptomatiques;
\item[$\bullet$] l'utilisation du masque facial peut r\'eduire la transmission par contact en emp\^echant les porteurs de se toucher la bouche ou le nez avec leurs mains ou d'autres objets potentiellement contamin\'es par le virus;
\item[$\bullet$] le port du masque est un moyen pour pr\'evenir la transmission des a\'erosols, qui peut \^etre \`a l'origine des cas de Covid-19.
\end{enumerate}
L'ampleur de ces effets suppl\'ementaires n'est pas totalement connue. En effet, si la transmission se fait principalement par contact alors la contribution de l'utilisation du masque pourrait \^etre faible par rapport aux mesures d'hygi\`enes (se laver les mains plus r\'eguli\`eremment ou \'eviter les contacts physiques et \'eviter les lieux publics bond\'es). 
La distanciation sociale peut largement emp\^echer la transmission par contact et gouttelettes mais est beaucoup moins efficace contre la transmission par a\'erosols. \\

\noindent Les simulations num\'eriques utilisant des donn\'ees d'incidence de la Guadeloupe et de la Martinique sugg\`erent donc un r\^ole b\'en\'efique pour l'adoption massive du port de masques, m\^eme peu efficaces, avec un avantage probablement plus important dans la classe des individus masqu\'es. En somme, le port du masque est ainsi utile contre la propagation de l'\'epid\'emie de Covid-19.

\addcontentsline{toc}{section}{D\'eclaration de conflits d'int\'er\^ets}

\section*{D\'eclaration de conflits d'int\'er\^ets}
\noindent L'auteur ne d\'eclare aucun conflit d'int\'er\^et. Les bailleurs de fonds n'ont jou\'e aucun r\^ole: dans la conception de l'\'etude; dans la collecte, l'analyse ou l'interpr\'etation des donn\'ees; dans la r\'edaction du manuscrit, ou dans la d\'ecision de publier les r\'esultats.

\appendix

\section{Annexe - Th\'eor\`eme \ref{th_EI}}\label{A_EMMP}
\noindent Pour tout $t\in\mathbb{R}_+$, nous posons:
$$
x(t)=
\left(
\begin{array}{rl}
S_G(t) \\
I_G(t) \\
R_G(t)  \\
\overline{S}_G(t) \\
\overline{I}_G(t) \\
\overline{R}_G(t)  \\
S_M(t) \\
I_M(t) \\
R_M(t)  \\
\overline{S}_M(t) \\
\overline{I}_M(t) \\
\overline{R}_M(t)  \\
\end{array}
\right)
\ \ \mbox{ et } \ \  
f(t, x(t))=
\left(
\begin{array}{rl}
\dfrac{dS_G}{dt} \medskip \\
\dfrac{dI_G}{dt} \medskip \\
\dfrac{dR_G}{dt} \medskip \\
\dfrac{d\overline{S}_G}{dt} \medskip \\
\dfrac{d\overline{I}_G}{dt} \medskip \\
\dfrac{d\overline{R}_G}{dt} \medskip \\
\dfrac{dS_M}{dt} \medskip \\
\dfrac{dI_M}{dt} \medskip \\
\dfrac{dR_M}{dt} \medskip \\
\dfrac{d\overline{S}_M}{dt} \medskip \\
\dfrac{d\overline{I}_M}{dt} \medskip \\
\dfrac{d\overline{R}_M}{dt} \medskip \\
\end{array}
\right) .
$$
Alors, le syst\`eme d'\'equations (\ref{mp2}) peut s'\'ecrire sous la forme:
\begin{eqnarray}\label{edo_E}
x'(t)=f(t, x(t)) .
\end{eqnarray}
Soit, arbitrairement, $(t_0; x_0)\in \mathbb{R}_+\times\mathbb{R}_+^{12}$ la conditions initiale de notre mod\`ele telle que la solution du syst\`eme (\ref{mp2}) v\'erifie cette condition initiale.  \\
Alors, le probl\`eme de Cauchy associ\'e \`a l'\'equation diff\'erentielle (\ref{edo_E}) est d\'efini par:
\begin{eqnarray}\label{pcm}
\left\{
\begin{array}{ll}
x'(t)=f(t, x(t)) \\
x(t_0)=x_0 \ \mbox{ donn\'e} ,  \\
\end{array}
\right.
\end{eqnarray}
o\`u 
$$
f(x(t))=
A(t)
\left(
\begin{array}{cc}
S_G(t)I_G(t) \\
S_M(t)I_M(t) \\
S_G(t)I_M(t) \\
S_M(t)I_G(t) \\
\overline{S}_G(t)\overline{I}_G(t) \\
\overline{S}_G(t)\overline{I}_M(t) \\
S_G(t)\overline{I}_G(t) \\
\overline{S}_G(t)I_G(t) \\
S_G(t)\overline{I}_M(t) \\
\overline{S}_G(t)I_M(t) \\
\overline{S}_M(t)\overline{I}_M(t) \\
S_M(t)\overline{I}_G(t) \\
S_M(t)\overline{I}_M(t) \\
\overline{S}_M(t)I_G(t) \\
\overline{S}_M(t)\overline{I}_G(t) \\
\overline{S}_M(t)I_M(t) \\
\end{array}
\right)
+
B(t)
\left(
\begin{array}{rl}
S_G(t) \\
I_G(t) \\
R_G(t)  \\
\overline{S}_G(t) \\
\overline{I}_G(t) \\
\overline{R}_G(t)  \\
S_M(t) \\
I_M(t) \\
R_M(t)  \\
\overline{S}_M(t) \\
\overline{I}_M(t) \\
\overline{R}_M(t)  \\
\end{array}
\right) , 
$$
avec \\
$$
A(t)=
\left[
\begin{array}{ccccccccccccccccc}
v_1 & v_2 & v_3 & v_4 & v_5 & v_6 & v_7 & v_8 & v_9 & v_{10} & v_{11} & v_{12} & v_{13} & v_{14} & v_{15} & v_{28} \\ 
\end{array}
\right]
$$
et
$$
B(t)=
\left[
\begin{array}{cccccccccccc}
v_{16} & v_{18} & v_{24} & v_{20} & v_{21} & v_{25} & v_{17} & v_{19} & v_{26} & v_{22} & v_{23} & v_{27} \\ 
\end{array}
\right], 
$$
puis \\
$
v_1=
\left(
\begin{array}{cc}
-\dfrac{\beta_G}{N_G(t)} \\
\dfrac{\beta_G}{N_G(t)} \\
0 \\
-(1-\varepsilon_i)\dfrac{\beta_G}{N_G(t)}  \\
0 \\
0 \\
0 \\
0 \\
0 \\
0 \\
0 \\
0 \\
\end{array}
\right)
$
, 
$
v_2=
\left(
\begin{array}{cc}
0 \\
0 \\
0 \\
0 \\
0 \\
0 \\
-\dfrac{\beta_M}{N_M(t)} \\
\dfrac{\beta_M}{N_M(t)} \\
0 \\
-(1-\varepsilon_i)\dfrac{\beta_M}{N_M(t)}  \\
0 \\
0 \\
\end{array}
\right)
$
, 
$
v_3=
\left(
\begin{array}{cc}
-\tau\varphi_2(t)\dfrac{n_{GM}}{N_G^\ast(t)}\dfrac{\beta_M}{N_M(t)} \\
\tau\varphi_2(t)\dfrac{1}{N_G^\ast(t)}\dfrac{\beta_M}{N_M(t)} \\
0 \\
0 \\
0 \\
0 \\
\tau\varphi_2(t)\dfrac{n_{GM}}{N_G^\ast(t)}\dfrac{\beta_M}{N_M(t)} \\
\tau\varphi_2(t)\dfrac{1}{N_G^\ast(t)}\dfrac{\beta_M}{N_M(t)} \\
0 \\
0 \\
0 \\
0 \\
\end{array}
\right)
$, 
\\
$
v_4=
\left(
\begin{array}{cc}
\tau\varphi_2(t)\dfrac{n_{MG}}{N_M^\ast(t)}\dfrac{\beta_G}{N_G(t)} \\
\tau\varphi_2(t)\dfrac{1}{N_M^\ast(t)}\dfrac{\beta_G}{N_G(t)} \\
0 \\
0 \\
0 \\
0 \\
-\tau\varphi_2(t)\dfrac{n_{MG}}{N_M^\ast(t)}\dfrac{\beta_G}{N_G(t)} \\
\tau\varphi_2(t)\dfrac{1}{N_M^\ast(t)}\dfrac{\beta_G}{N_G(t)} \\
0 \\
0 \\
0 \\
0 \\
\end{array}
\right)
$
, 
$
v_5=
\left(
\begin{array}{cc}
0 \\
0 \\
0 \\
-(1-\varepsilon_i)(1-\varepsilon_e)\dfrac{\beta_G}{N_G(t)} \\
0 \\
0 \\
0 \\
0 \\
0 \\
0 \\
0 \\
0 \\
\end{array}
\right)
$
, 
$
v_6=
\left(
\begin{array}{cc}
0 \\
0 \\
0 \\
-(1-\varepsilon_i)(1-\varepsilon_e)\tau\varphi_2(t)\dfrac{n_{GM}}{N_G^\ast(t)} \\
0 \\
0 \\
0 \\
0 \\
0 \\
-(1-\varepsilon_i)(1-\varepsilon_e)\tau\varphi_2(t)\dfrac{n_{GM}}{N_G^\ast(t)}\dfrac{\beta_M}{N_M(t)} \\
0 \\
0 \\
\end{array}
\right)
$
\\
$
v_7=
\left(
\begin{array}{cc}
-(1-\varepsilon_e)\dfrac{\beta_G}{N_G(t)} \\
0 \\
0 \\
0 \\
0 \\
0 \\
0 \\
0 \\
0 \\
0 \\
0 \\
0 \\
\end{array}
\right)
$
, 
$
v_8=
\left(
\begin{array}{cc}
0 \\
0 \\
0 \\
0 \\
\dfrac{\beta_G}{N_G(t)} \\
0 \\
0 \\
0 \\
0 \\
0 \\
0 \\
0 \\
\end{array}
\right)
$
,   
$
v_9=
\left(
\begin{array}{cc}
-(1-\varepsilon_e)\tau\varphi_2(t)\dfrac{n_{GM}}{N_G^\ast(t)}\dfrac{\beta_M}{N_M(t)} \\
0 \\
0 \\
0 \\
0 \\
0 \\
(1-\varepsilon_e)\tau\varphi_2(t)\dfrac{n_{GM}}{N_G^\ast(t)}\dfrac{\beta_M}{N_M(t)} \\
0 \\
0 \\
0 \\
0 \\
0 \\
\end{array}
\right)
$
, 
\\
$
v_{10}=
\left(
\begin{array}{cc}
0 \\
0 \\
0 \\
-(1-\varepsilon_i)\tau\varphi_2(t)\dfrac{n_{GM}}{N_G^\ast(t)}\dfrac{\beta_M}{N_M(t)} \\
\tau\varphi_2(t)\dfrac{1}{N_G^\ast(t)}\dfrac{\beta_M}{N_M(t)} \\
0 \\
0 \\
0 \\
0 \\
(1-\varepsilon_i)\tau\varphi_2(t)\dfrac{n_{GM}}{N_G^\ast(t)}\dfrac{\beta_M}{N_M(t)}  \\
\tau\varphi_2(t)\dfrac{1}{N_G^\ast(t)}\dfrac{\beta_M}{N_M(t)}  \\
0 \\
\end{array}
\right)
$
,     
$
v_{11}=
\left(
\begin{array}{cc}
0 \\
0 \\
0 \\
0 \\
0 \\
0 \\
0 \\
0 \\
0 \\
-(1-\varepsilon_i)(1-\varepsilon_e)\dfrac{\beta_M}{N_M(t)}  \\
0  \\
0 \\
\end{array}
\right)
$
, 
$
v_{12}=
\left(
\begin{array}{cc}
(1-\varepsilon_e)\tau\varphi_2(t)\dfrac{n_{MG}}{N_M^\ast(t)}\dfrac{\beta_G}{N_G(t)}  \\
0 \\
0 \\
0 \\
0 \\
0 \\
-(1-\varepsilon_e)\tau\varphi_2(t)\dfrac{n_{MG}}{N_M^\ast(t)}\dfrac{\beta_G}{N_G(t)}  \\
0 \\
0 \\
0 \\
0 \\
0 \\
\end{array}
\right)
$ 
\\
$
v_{13}=
\left(
\begin{array}{cc}
0  \\
0 \\
0 \\
0 \\
0 \\
0 \\
(1-\varepsilon_e)\dfrac{\beta_M}{N_M(t)}  \\
0 \\
0 \\
0 \\
0 \\
0 \\
\end{array}
\right)
$
, 
$
v_{14}=
\left(
\begin{array}{cc}
0  \\
0 \\
0 \\
(1-\varepsilon_i)\tau\varphi_2(t)\dfrac{n_{MG}}{N_M^\ast(t)}\dfrac{\beta_G}{N_G(t)}  \\
\tau\varphi_2(t)\dfrac{1}{N_M^\ast(t)}\dfrac{\beta_G}{N_G(t)}  \\
0 \\
0 \\
0 \\
0 \\
-(1-\varepsilon_i)\tau\varphi_2(t)\dfrac{n_{MG}}{N_M^\ast(t)}\dfrac{\beta_G}{N_G(t)}  \\
\tau\varphi_2(t)\dfrac{1}{N_M^\ast(t)}\dfrac{\beta_G}{N_G(t)}  \\
0 \\
\end{array}
\right)
$
,  
$
v_{15}=
\left(
\begin{array}{cc}
0  \\
0 \\
0 \\
(1-\varepsilon_i)(1-\varepsilon_e)\tau\varphi_2(t)\dfrac{n_{MG}}{N_M^\ast(t)}\dfrac{\beta_G}{N_G(t)}  \\
0 \\
0 \\
0 \\
0 \\
0 \\
-(1-\varepsilon_i)(1-\varepsilon_e)\tau\varphi_2(t)\dfrac{n_{MG}}{N_M^\ast(t)}\dfrac{\beta_G}{N_G(t)}  \\
0 \\
0 \\
\end{array}
\right)
$
\\
$
v_{16}=
\left(
\begin{array}{cc}
(2-\varepsilon_e)\varphi(t)\dfrac{n_{GM}}{N_G^\ast(t)} \\
0  \\
0 \\
0 \\
0 \\
0 \\
-(2-\varepsilon_e)\varphi(t)\dfrac{n_{GM}}{N_G^\ast(t)} \\
0 \\
0 \\
0 \\
0 \\
0 \\
\end{array}
\right)
$
, 
$
v_{17}=
\left(
\begin{array}{cc}
-(2-\varepsilon_e)\varphi(t)\dfrac{n_{MG}}{N_M^\ast(t)} \\
0  \\
0 \\
0 \\
0 \\
0 \\
(2-\varepsilon_e)\varphi(t)\dfrac{n_{MG}}{N_M^\ast(t)} \\
0 \\
0 \\
0 \\
0 \\
0 \\
\end{array}
\right)
$
, 
$
v_{18}=
\left(
\begin{array}{cc}
0 \\
-\gamma+\eta\varphi(t)\dfrac{n_{GM}}{N_G^\ast(t)} \\
\gamma  \\
0 \\
0 \\
0 \\
0 \\
0 \\
0 \\
0 \\
0 \\
0 \\
\end{array}
\right)
$,
\\
$
v_{19}=
\left(
\begin{array}{cc}
0 \\
0 \\
0 \\
0 \\
0 \\
0 \\
0 \\
-\gamma+\eta\varphi(t)\dfrac{n_{MG}}{N_M^\ast(t)} \\
\gamma \\
0 \\
0 \\
0 \\
\end{array}
\right)
$
, 
$
v_{20}=
\left(
\begin{array}{cc}
0 \\
0 \\
0 \\
(1-\varepsilon_i)(2-\varepsilon_e)\varphi(t)\dfrac{n_{GM}}{N_G^\ast(t)} \\
0 \\
0 \\
0 \\
0 \\
0 \\
-(1-\varepsilon_i)(2-\varepsilon_e)\varphi(t)\dfrac{n_{GM}}{N_G^\ast(t)} \\
0 \\
0 \\
\end{array}
\right)
$
, 
$
v_{21}=
\left(
\begin{array}{cc}
0 \\
0 \\
0 \\
0 \\
-\gamma+\eta\varphi(t)\dfrac{n_{GM}}{N_G^\ast(t)} \\
\gamma \\
0 \\
0 \\
0 \\
0 \\
-\eta\varphi(t)\dfrac{n_{GM}}{N_G^\ast(t)} \\
0 \\
\end{array}
\right)
$, 
\\
$
v_{22}=
\left(
\begin{array}{cc}
0 \\
0 \\
0 \\
-(1-\varepsilon_i)(2-\varepsilon_e)\varphi(t)\dfrac{n_{MG}}{N_M^\ast(t)} \\
0 \\
0 \\
0 \\
0 \\
0 \\
(1-\varepsilon_i)(2-\varepsilon_e)\varphi(t)\dfrac{n_{MG}}{N_M^\ast(t)} \\
0 \\
0 \\
\end{array}
\right)
$
, 
$
v_{23}=
\left(
\begin{array}{cc}
0 \\
0 \\
0 \\
0 \\
-\eta\varphi(t)\dfrac{n_{MG}}{N_M^\ast(t)} \\
0 \\
0 \\
0 \\
0 \\
0 \\
-\gamma+\eta\varphi(t)\dfrac{n_{MG}}{N_M^\ast(t)} \\
\gamma \\
\end{array}
\right)
$
, 
$
v_{24}=
\left(
\begin{array}{cc}
0 \\
0 \\
\varphi(t)\dfrac{n_{GM}}{N_G^\ast(t)} \\
0 \\
0 \\
0 \\
0 \\
0 \\
-\varphi(t)\dfrac{n_{GM}}{N_G^\ast(t)} \\
0 \\
0 \\
0 \\
\end{array}
\right)
$,  
\\
$
v_{25}=
\left(
\begin{array}{cc}
0 \\
0 \\
0 \\
0 \\
0 \\
\varphi(t)\dfrac{n_{GM}}{N_G^\ast(t)} \\
0 \\
0 \\
0 \\
0 \\
0 \\
-\varphi(t)\dfrac{n_{GM}}{N_G^\ast(t)} \\
\end{array}
\right)
$
, 
$
v_{26}=
\left(
\begin{array}{cc}
0 \\
0 \\
-\varphi(t)\dfrac{n_{MG}}{N_M^\ast(t)} \\
0 \\
0 \\
0 \\
0 \\
0 \\
\varphi(t)\dfrac{n_{MG}}{N_M^\ast(t)} \\
0 \\
0 \\
0 \\
\end{array}
\right)
$
, 
$
v_{27}=
\left(
\begin{array}{cc}
0 \\
0 \\
0 \\
0 \\
0 \\
-\varphi(t)\dfrac{n_{MG}}{N_M^\ast(t)} \\
0 \\
0 \\
0 \\
0 \\
0 \\
\varphi(t)\dfrac{n_{MG}}{N_M^\ast(t)} \\
\end{array}
\right)
$ 
et 
$
v_{28}=
\left(
\begin{array}{cc}
0 \\
0 \\
0 \\
0 \\
0 \\
0 \\
0 \\
0 \\
0 \\
0 \\
\dfrac{\beta_M}{N_M(t)} \\
0 \\
\end{array}
\right)
$. \\

\noindent Par ailleurs, $(S_G(t), I_G(t), R_G(t), \overline{S}_G(t), \overline{I}_G(t), \overline{R}_G(t), S_M(t), I_M(t), R_M(t), \overline{S}_M(t), \overline{I}_M(t), \overline{R}_M(t))$ est une base de $\mathbb{R}_+^{12}$ et $x(t)$ d\'esigne une matrice-colonne des coordonn\'ees dans cette base. Si, pour tout $t\in\mathbb{R}_+$, on pose:
$$
\Psi(t, x(t))=
A(t)
\left(
\begin{array}{cc}
S_G(t)I_G(t) \\
S_M(t)I_M(t) \\
S_G(t)I_M(t) \\
S_M(t)I_G(t) \\
\overline{S}_G(t)\overline{I}_G(t) \\
\overline{S}_G(t)\overline{I}_M(t) \\
S_G(t)\overline{I}_G(t) \\
\overline{S}_G(t)I_G(t) \\
S_G(t)\overline{I}_M(t) \\
\overline{S}_G(t)I_M(t) \\
\overline{S}_M(t)\overline{I}_M(t) \\
S_M(t)\overline{I}_G(t) \\
S_M(t)\overline{I}_M(t) \\
\overline{S}_M(t)I_G(t) \\
\overline{S}_M(t)\overline{I}_G(t) \\
\overline{S}_M(t)I_M(t) \\
\end{array}
\right) ,
$$
alors $\Psi$ une forme bilin\'eaire sym\'etrique d\'efinie sur $\mathbb{R}^{12}$ et, pour tout $t\in\mathbb{R}_+$, on a:
$$
\Psi(t, x(t))=
\left(
\begin{array}{cccccccccccc}
~^tx(t)A_1(t)x(t) \\
~^tx(t)A_2(t)x(t)  \\
~^tx(t)A_3(t)x(t) \\
~^tx(t)A_4(t)x(t)  \\
~^tx(t)A_5(t)x(t)  \\
~^tx(t)A_6(t)x(t) \\
~^tx(t)A_7(t)x(t)  \\
~^tx(t)A_8(t)x(t)  \\
~^tx(t)A_9(t)x(t)  \\
~^tx(t)A_{10}(t)x(t) \\
~^tx(t)A_{11}(t)x(t) \\
~^tx(t)A_{12}(t)x(t) \\
\end{array}
\right) ,
$$
o\`u $~^tx(t)$ d\'esigne la transpos\'ee de la matrice-colonne $x(t)$ et 
$$
A_1 = \left(
\begin{array}{cccccccccccc}
0 & 0 & 0 & 0 & -\dfrac{(1-\varepsilon_e)\beta_G}{N_G(t)} & 0 & 0 & -\tau\varphi_2(t)\dfrac{n_{GM}\beta_M}{N_G^\ast(t)N_M(t)} & 0 & 0 & a_{1,11}^{(1)} & 0  \\
\dfrac{-\beta_G}{N_G(t)} & 0 & 0 & 0 & 0 & 0 & 0 & 0 & 0 & 0 & 0 & 0  \\
0 & 0 & 0 & 0 & 0 & 0 & 0 & 0 & 0 & 0 & 0 & 0  \\
0 & 0 & 0 & 0 & 0 & 0 & 0 & 0 & 0 & 0 & 0 & 0  \\
0 & 0 & 0 & 0 & 0 & 0 & a_{5,7}^{(1)} & 0 & 0 & 0 & 0 & 0  \\
0 & 0 & 0 & 0 & 0 & 0 & 0 & 0 & 0 & 0 & 0 & 0  \\
0 & \tau\varphi_2(t)\dfrac{n_{MG}}{N_M^\ast(t)}\dfrac{\beta_G}{N_G(t)} & 0 & 0 & 0 & 0 & 0 & 0 & 0 & 0 & 0 & 0  \\
0 & 0 & 0 & 0 & 0 & 0 & 0 & 0 & 0 & 0 & 0 & 0   \\
0 & 0 & 0 & 0 & 0 & 0 & 0 & 0 & 0 & 0 & 0 & 0  \\
0 & 0 & 0 & 0 & 0 & 0 & 0 & 0 & 0 & 0 & 0 & 0  \\
0 & 0 & 0 & 0 & 0 & 0 & 0 & 0 & 0 & 0 & 0 & 0  \\
0 & 0 & 0 & 0 & 0 & 0 & 0 & 0 & 0 & 0 & 0 & 0  \\
\end{array}
\right)
$$
avec 
$$
\begin{tabular}{ll}
$a_{1,11}^{(1)}=-(1-\varepsilon_e)\tau\varphi_2(t)\dfrac{n_{GM}}{N_G^\ast(t)}\dfrac{\beta_M}{N_M(t)}$, \qquad $a_{5,7}^{(1)}=(1-\varepsilon_e)\tau\varphi_2(t)\dfrac{n_{MG}}{N_M^\ast(t)}\dfrac{\beta_G}{N_G(t)}$,
\end{tabular}
$$

$$
A_2(t) = \left(
\begin{array}{cccccccccccc}
0 & 0 & 0 & 0 & 0 & 0 & 0 & \tau\varphi_2(t)\dfrac{1}{N_G^\ast(t)}\dfrac{\beta_M}{N_M(t)} & 0 & 0 & 0 & 0  \\
\dfrac{\beta_G}{N_G(t)} & 0 & 0 & 0 & 0 & 0 & 0 & 0 & 0 & 0 & 0 & 0  \\
0 & 0 & 0 & 0 & 0 & 0 & 0 & 0 & 0 & 0 & 0 & 0  \\
0 & 0 & 0 & 0 & 0 & 0 & 0 & 0 & 0 & 0 & 0 & 0  \\
0 & 0 & 0 & 0 & 0 & 0 & 0 & 0 & 0 & 0 & 0 & 0  \\
0 & 0 & 0 & 0 & 0 & 0 & 0 & 0 & 0 & 0 & 0 & 0  \\
0 & \tau\varphi_2(t)\dfrac{1}{N_M^\ast(t)}\dfrac{\beta_G}{N_G(t)} & 0 & 0 & 0 & 0 & 0 & 0 & 0 & 0 & 0 & 0  \\
0 & 0 & 0 & 0 & 0 & 0 & 0 & 0 & 0 & 0 & 0 & 0   \\
0 & 0 & 0 & 0 & 0 & 0 & 0 & 0 & 0 & 0 & 0 & 0  \\
0 & 0 & 0 & 0 & 0 & 0 & 0 & 0 & 0 & 0 & 0 & 0  \\
0 & 0 & 0 & 0 & 0 & 0 & 0 & 0 & 0 & 0 & 0 & 0  \\
0 & 0 & 0 & 0 & 0 & 0 & 0 & 0 & 0 & 0 & 0 & 0  \\
\end{array}
\right),
$$

$$
A_4 = \left(
\begin{array}{cccccccccccc}
0 & 0 & 0 & 0 & 0 & 0 & 0 & 0 & 0 & 0 & 0 & 0  \\
-\dfrac{(1-\varepsilon_i)\beta_G}{N_G(t)} & 0 & 0 & 0 & 0 & 0 & 0 & 0 & 0 & a_{2,10}^{(4)} & 0 & 0  \\
0 & 0 & 0 & 0 & 0 & 0 & 0 & 0 & 0 & 0 & 0 & 0  \\
0 & 0 & 0 & 0 & 0 & 0 & 0 & -(1-\varepsilon_i)\tau\varphi_2(t)\dfrac{n_{GM}}{N_G^\ast(t)}\dfrac{\beta_M}{N_M(t)} & 0 & 0 & a_{4,11}^{(4)} & 0  \\
0 & 0 & 0 & a_{5,4}^{(4)} & 0 & 0 & 0 & 0 & 0 & 0 & 0 & 0  \\
0 & 0 & 0 & 0 & 0 & 0 & 0 & 0 & 0 & 0 & 0 & 0  \\
0 & 0 & 0 & 0 & 0 & 0 & 0 & 0 & 0 & 0 & 0 & 0  \\
0 & 0 & 0 & 0 & 0 & 0 & 0 & 0 & 0 & 0 & 0 & 0   \\
0 & 0 & 0 & 0 & 0 & 0 & 0 & 0 & 0 & 0 & 0 & 0  \\
0 & 0 & 0 & 0 & a_{10,5}^{(4)} & 0 & 0 & 0 & 0 & 0 & 0 & 0  \\
0 & 0 & 0 & 0 & 0 & 0 & 0 & 0 & 0 & 0 & 0 & 0  \\
0 & 0 & 0 & 0 & 0 & 0 & 0 & 0 & 0 & 0 & 0 & 0  \\
\end{array}
\right),
$$
avec
$$
\begin{tabular}{lll}
$a_{2,10}^{(4)}=(1-\varepsilon_i)\tau\varphi_2(t)\dfrac{n_{MG}}{N_M^\ast(t)}\dfrac{\beta_G}{N_G(t)}$, & \qquad $a_{5,4}^{(4)}=-(1-\varepsilon_i)(1-\varepsilon_e)\dfrac{\beta_G}{N_G(t)}$,  \medskip \\
$a_{4,11}^{(4)}=-(1-\varepsilon_i)(1-\varepsilon_e)\tau\varphi_2(t)\dfrac{n_{GM}}{N_G^\ast(t)}$, & \qquad $a_{10,5}^{(4)}=(1-\varepsilon_i)(1-\varepsilon_e)\tau\varphi_2(t)\dfrac{n_{MG}}{N_M^\ast(t)}\dfrac{\beta_G}{N_G(t)}$, \medskip \\
\end{tabular}
$$
$$
A_5(t) = \left(
\begin{array}{cccccccccccc}
0 & 0 & 0 & 0 & 0 & 0 & 0 & 0 & 0 & 0 & 0 & 0  \\
0 & 0 & 0 & 0 & 0 & 0 & 0 & 0 & 0 & \tau\varphi_2(t)\dfrac{1}{N_M^\ast(t)}\dfrac{\beta_G}{N_G(t)} & 0 & 0  \\
0 & 0 & 0 & 0 & 0 & 0 & 0 & 0 & 0 & 0 & 0 & 0  \\
0 & \dfrac{\beta_G}{N_G(t)} & 0 & 0 & 0 & 0 & 0 & \tau\varphi_2(t)\dfrac{1}{N_G^\ast(t)}\dfrac{\beta_M}{N_M(t)} & 0 & 0 & 0 & 0  \\
0 & 0 & 0 & 0 & 0 & 0 & 0 & 0 & 0 & 0 & 0 & 0  \\
0 & 0 & 0 & 0 & 0 & 0 & 0 & 0 & 0 & 0 & 0 & 0  \\
0 & 0 & 0 & 0 & 0 & 0 & 0 & 0 & 0 & 0 & 0 & 0  \\
0 & 0 & 0 & 0 & 0 & 0 & 0 & 0 & 0 & 0 & 0 & 0   \\
0 & 0 & 0 & 0 & 0 & 0 & 0 & 0 & 0 & 0 & 0 & 0  \\
0 & 0 & 0 & 0 & 0 & 0 & 0 & 0 & 0 & 0 & 0 & 0  \\
0 & 0 & 0 & 0 & 0 & 0 & 0 & 0 & 0 & 0 & 0 & 0  \\
0 & 0 & 0 & 0 & 0 & 0 & 0 & 0 & 0 & 0 & 0 & 0  \\
\end{array}
\right),
$$

$$
A_7(t) = \left(
\begin{array}{cccccccccccc}
0 & 0 & 0 & 0 & 0 & 0 & 0 & 0 & 0 & 0 & a_{1,11}^{(7)} & 0  \\
0 & 0 & 0 & 0 & 0 & 0 & 0 & 0 & 0 & 0 & 0 & 0  \\
0 & 0 & 0 & 0 & 0 & 0 & 0 & 0 & 0 & 0 & 0 & 0  \\
0 & 0 & 0 & 0 & 0 & 0 & 0 & 0 & 0 & 0 & 0 & 0  \\
0 & 0 & 0 & 0 & 0 & 0 & 0 & 0 & 0 & 0 & 0 & 0  \\
0 & 0 & 0 & 0 & 0 & 0 & 0 & 0 & 0 & 0 & 0 & 0  \\
0 & -\tau\varphi_2(t)\dfrac{n_{MG}}{N_M^\ast(t)}\dfrac{\beta_G}{N_G(t)} & 0 & 0 & a_{7,5}^{(7)} & 0 & 0 & \dfrac{-\beta_M}{N_M(t)} & 0 & 0 & 0 & 0  \\
a_{8,1}^{(7)} & 0 & 0 & 0 & 0 & 0 & 0 & 0 & 0 & 0 & 0 & 0   \\
0 & 0 & 0 & 0 & 0 & 0 & 0 & 0 & 0 & 0 & 0 & 0  \\
0 & 0 & 0 & 0 & 0 & 0 & 0 & 0 & 0 & 0 & 0 & 0  \\
0 & 0 & 0 & 0 & 0 & 0 & (1-\varepsilon_e)\dfrac{\beta_M}{N_M(t)} & 0 & 0 & 0 & 0 & 0  \\
0 & 0 & 0 & 0 & 0 & 0 & 0 & 0 & 0 & 0 & 0 & 0  \\
\end{array}
\right),
$$
avec
$$
\begin{tabular}{lll}
$a_{1,11}^{(7)}=(1-\varepsilon_e)\tau\varphi_2(t)\dfrac{n_{GM}}{N_G^\ast(t)}\dfrac{\beta_M}{N_M(t)}$, \medskip $a_{7,5}^{(7)}=-(1-\varepsilon_e)\tau\varphi_2(t)\dfrac{n_{MG}}{N_M^\ast(t)}\dfrac{\beta_G}{N_G(t)}$, \medskip $a_{8,1}^{(7)}=\tau\varphi_2(t)\dfrac{n_{GM}}{N_G^\ast(t)}\dfrac{\beta_M}{N_M(t)}$,
\end{tabular}
$$

$$
A_8(t) = \left(
\begin{array}{cccccccccccc}
0 & 0 & 0 & 0 & 0 & 0 & 0 & 0 & 0 & 0 & 0 & 0  \\
0 & 0 & 0 & 0 & 0 & 0 & \tau\varphi_2(t)\dfrac{1}{N_M^\ast(t)}\dfrac{\beta_G}{N_G(t)} & 0 & 0 & 0 & 0 & 0  \\
0 & 0 & 0 & 0 & 0 & 0 & 0 & 0 & 0 & 0 & 0 & 0  \\
0 & 0 & 0 & 0 & 0 & 0 & 0 & 0 & 0 & 0 & 0 & 0  \\
0 & 0 & 0 & 0 & 0 & 0 & 0 & 0 & 0 & 0 & 0 & 0  \\
0 & 0 & 0 & 0 & 0 & 0 & 0 & 0 & 0 & 0 & 0 & 0  \\
0 & 0 & 0 & 0 & 0 & 0 & 0 & \dfrac{\beta_M}{N_M(t)} & 0 & 0 & 0 & 0  \\
\tau\varphi_2(t)\dfrac{1}{N_G^\ast(t)}\dfrac{\beta_M}{N_M(t)} & 0 & 0 & 0 & 0 & 0 & 0 & 0 & 0 & 0 & 0 & 0   \\
0 & 0 & 0 & 0 & 0 & 0 & 0 & 0 & 0 & 0 & 0 & 0  \\
0 & 0 & 0 & 0 & 0 & 0 & 0 & 0 & 0 & 0 & 0 & 0  \\
0 & 0 & 0 & 0 & 0 & 0 & 0 & 0 & 0 & 0 & 0 & 0  \\
0 & 0 & 0 & 0 & 0 & 0 & 0 & 0 & 0 & 0 & 0 & 0  \\
\end{array}
\right),
$$

$$
A_{10} = \left(
\begin{array}{cccccccccccc}
0 & 0 & 0 & 0 & 0 & 0 & 0 & 0 & 0 & 0 & 0 & 0  \\
0 & 0 & 0 & 0 & 0 & 0 & 0 & 0 & 0 & 0 & 0 & 0  \\
0 & 0 & 0 & 0 & 0 & 0 & 0 & 0 & 0 & 0 & 0 & 0  \\
0 & 0 & 0 & 0 & 0 & 0 & 0 & 0 & 0 & 0 & a_{4,11}^{(10)} & 0  \\
0 & 0 & 0 & 0 & 0 & 0 & 0 & 0 & 0 & 0 & 0 & 0  \\
0 & 0 & 0 & 0 & 0 & 0 & 0 & 0 & 0 & 0 & 0 & 0  \\
0 & 0 & 0 & 0 & 0 & 0 & 0 & -(1-\varepsilon_i)\dfrac{\beta_M}{N_M(t)} & 0 & 0 & 0 & 0  \\
0 & 0 & 0 & (1-\varepsilon_i)\tau\varphi_2(t)\dfrac{n_{GM}}{N_G^\ast(t)}\dfrac{\beta_M}{N_M(t)} & 0 & 0 & 0 & 0 & 0 & 0 & 0 & 0   \\
0 & 0 & 0 & 0 & 0 & 0 & 0 & 0 & 0 & 0 & 0 & 0  \\
0 & a_{10,2}^{(10)} & 0 & 0 & a_{10,5}^{(10)} & 0 & 0 & 0 & 0 & 0 & 0 & 0  \\
0 & 0 & 0 & 0 & 0 & 0 & 0 & 0 & 0 & a_{11,10}^{(10)} & 0 & 0  \\
0 & 0 & 0 & 0 & 0 & 0 & 0 & 0 & 0 & 0 & 0 & 0  \\
\end{array}
\right),
$$
avec
$$
\begin{tabular}{lll}
$a_{4,11}^{(10)}=-(1-\varepsilon_i)(1-\varepsilon_e)\tau\varphi_2(t)\dfrac{n_{GM}}{N_G^\ast(t)}\dfrac{\beta_M}{N_M(t)}$, & \qquad $a_{10,2}^{(10)}=-(1-\varepsilon_i)\tau\varphi_2(t)\dfrac{n_{MG}}{N_M^\ast(t)}\dfrac{\beta_G}{N_G(t)}$, \medskip \\ 
$a_{10,5}^{(10)}=-(1-\varepsilon_i)(1-\varepsilon_e)\tau\varphi_2(t)\dfrac{n_{MG}}{N_M^\ast(t)}\dfrac{\beta_G}{N_G(t)}$, & \qquad $a_{11,10}^{(10)}=-(1-\varepsilon_i)(1-\varepsilon_e)\dfrac{\beta_M}{N_M(t)}$, \medskip \\
\end{tabular}
$$
$$
A_{11} = \left(
\begin{array}{cccccccccccc}
0 & 0 & 0 & 0 & 0 & 0 & 0 & 0 & 0 & 0 & 0 & 0  \\
0 & 0 & 0 & 0 & 0 & 0 & 0 & 0 & 0 & 0 & 0 & 0  \\
0 & 0 & 0 & 0 & 0 & 0 & 0 & 0 & 0 & 0 & 0 & 0  \\
0 & 0 & 0 & 0 & 0 & 0 & 0 & 0 & 0 & 0 & 0 & 0  \\
0 & 0 & 0 & 0 & 0 & 0 & 0 & 0 & 0 & 0 & 0 & 0  \\
0 & 0 & 0 & 0 & 0 & 0 & 0 & 0 & 0 & 0 & 0 & 0  \\
0 & 0 & 0 & 0 & 0 & 0 & 0 & 0 & 0 & 0 & 0 & 0  \\
0 & 0 & 0 & -(1-\varepsilon_i)(1-\varepsilon_e)\tau\varphi_2(t)\dfrac{n_{GM}}{N_G^\ast(t)}\dfrac{\beta_M}{N_M(t)} & 0 & 0 & 0 & 0 & 0 & 0 & 0 & 0   \\
0 & 0 & 0 & 0 & 0 & 0 & 0 & 0 & 0 & 0 & 0 & 0  \\
0 & \tau\varphi_2(t)\dfrac{1}{N_M^\ast(t)}\dfrac{\beta_G}{N_G(t)} & 0 & 0 & 0 & 0 & 0 & \dfrac{\beta_M}{N_M(t)} & 0 & 0 & 0 & 0  \\
0 & 0 & 0 & 0 & 0 & 0 & 0 & 0 & 0 & 0 & 0 & 0  \\
0 & 0 & 0 & 0 & 0 & 0 & 0 & 0 & 0 & 0 & 0 & 0  \\
\end{array}
\right),
$$
et
$$
A_3(t)=A_6(t)=A_9(t)=A_{12}(t) = \left(
\begin{array}{cccccccccccc}
0 & 0 & 0 & 0 & 0 & 0 & 0 & 0 & 0 & 0 & 0 & 0  \\
0 & 0 & 0 & 0 & 0 & 0 & 0 & 0 & 0 & 0 & 0 & 0  \\
0 & 0 & 0 & 0 & 0 & 0 & 0 & 0 & 0 & 0 & 0 & 0  \\
0 & 0 & 0 & 0 & 0 & 0 & 0 & 0 & 0 & 0 & 0 & 0  \\
0 & 0 & 0 & 0 & 0 & 0 & 0 & 0 & 0 & 0 & 0 & 0  \\
0 & 0 & 0 & 0 & 0 & 0 & 0 & 0 & 0 & 0 & 0 & 0  \\
0 & 0 & 0 & 0 & 0 & 0 & 0 & 0 & 0 & 0 & 0 & 0  \\
0 & 0 & 0 & 0 & 0 & 0 & 0 & 0 & 0 & 0 & 0 & 0   \\
0 & 0 & 0 & 0 & 0 & 0 & 0 & 0 & 0 & 0 & 0 & 0  \\
0 & 0 & 0 & 0 & 0 & 0 & 0 & 0 & 0 & 0 & 0 & 0  \\
0 & 0 & 0 & 0 & 0 & 0 & 0 & 0 & 0 & 0 & 0 & 0  \\
0 & 0 & 0 & 0 & 0 & 0 & 0 & 0 & 0 & 0 & 0 & 0  \\
\end{array}
\right).
$$ 
Ainsi, pour tout $t\in\mathbb{R}_+$, on a:
$$
f(t, x(t))=\Psi(t, x(t))+B(t)x(t) ,
$$ 
o\`u 
$$
B(t)= \left(
\begin{array}{cccccccccccc}
b_{1,1} & 0 & 0 & 0 & 0 & 0 & b_{1,7} & 0 & 0 & 0 & 0 & 0  \\
0 & b_{2,2} & 0 & 0 & 0 & 0 & 0 & 0 & 0 & 0 & 0 & 0  \\
0 & \gamma & \varphi(t)\dfrac{n_{GM}}{N^\ast_G(t)} & 0 & 0 & 0 & 0 & 0 & -\varphi(t)\dfrac{n_{MG}}{N^\ast_M(t)} & 0 & 0 & 0  \\
0 & 0 & 0 &  b_{4,4} & 0 & 0 & 0 & 0 & 0 & b_{4,10} & 0 & 0  \\
0 & 0 & 0 & 0 & b_{5,5} & 0 & 0 & 0 & 0 & 0 & b_{5,11} & 0  \\
0 & 0 & 0 & 0 & \gamma & b_{6,6} & 0 & 0 & 0 & 0 & 0 & b_{6,12}  \\
b_{7,1} & 0 & 0 & 0 & 0 & 0 & b_{7,7} & 0 & 0 & 0 & 0 & 0  \\
0 & 0 & 0 & 0 & 0 & 0 & 0 & b_{8,8} & 0 & 0 & 0 & 0   \\
0 & 0 & -\varphi(t)\dfrac{n_{GM}}{N^\ast_G(t)} & 0 & 0 & 0 & 0 & \gamma & \varphi(t)\dfrac{n_{MG}}{N^\ast_M(t)} & 0 & 0 & 0  \\
0 & 0 & 0 & b_{10,4} & 0 & 0 & 0 & 0 & 0 & b_{10,10} & 0 & 0  \\
0 & 0 & 0 & 0 & b_{11,5} & 0 & 0 & 0 & 0 & 0 & b_{11,11} & 0  \\
0 & 0 & 0 & 0 & 0 & b_{12,6} & 0 & 0 & 0 & 0 & \gamma & b_{12,12}  \\
\end{array}
\right)
$$ 
avec
$$
\begin{tabular}{lll}
$b_{1,1}=(2-\varepsilon_e)\varphi(t)\dfrac{n_{GM}}{N_G^\ast(t)}$, & \qquad $b_{7,1}=-(2-\varepsilon_e)\varphi(t)\dfrac{n_{GM}}{N_G^\ast(t)}$, \medskip \\ 
$b_{2,2}=-\gamma+\eta\varphi(t)\dfrac{n_{GM}}{N_G^\ast(t)}$, & \qquad $b_{8,8}=-\gamma+\eta\varphi(t)\dfrac{n_{MG}}{N_M^\ast(t)}$, & \qquad  \medskip \\
$b_{4,4}=(1-\varepsilon_i)(2-\varepsilon_e)\varphi(t)\dfrac{n_{GM}}{N_G^\ast(t)}$, & \qquad $b_{10,4}=-(1-\varepsilon_i)(2-\varepsilon_e)\varphi(t)\dfrac{n_{GM}}{N_G^\ast(t)}$, \medskip \\
$b_{5,5}=-\gamma+\eta\varphi(t)\dfrac{n_{GM}}{N_G^\ast(t)}$, & \qquad $b_{11,5}=-\eta\varphi(t)\dfrac{n_{GM}}{N_G^\ast(t)}$, \medskip \\
$b_{6,6}=\varphi(t)\dfrac{n_{GM}}{N^\ast_G(t)}$, & \qquad $b_{6,12}=-\varphi(t)\dfrac{n_{MG}}{N^\ast_M(t)}$, \medskip \\ 
$b_{1,7}=-(2-\varepsilon_e)\varphi(t)\dfrac{n_{MG}}{N_M^\ast(t)}$, & \qquad $b_{7,7}=(2-\varepsilon_e)\varphi(t)\dfrac{n_{MG}}{N_M^\ast(t)}$, \medskip \\
$b_{4,10}=-(1-\varepsilon_i)(2-\varepsilon_e)\varphi(t)\dfrac{n_{MG}}{N_M^\ast(t)}$, & \qquad $b_{10,10}=(1-\varepsilon_i)(2-\varepsilon_e)\varphi(t)\dfrac{n_{MG}}{N_M^\ast(t)}$, \medskip \\
$b_{5,11}=-\eta\varphi(t)\dfrac{n_{MG}}{N_M^\ast(t)}$, & \qquad $b_{11,11}=-\gamma+\eta\varphi(t)\dfrac{n_{MG}}{N_M^\ast(t)}$, \medskip \\
$b_{12,6}=-\varphi(t)\dfrac{n_{GM}}{N^\ast_G(t)}$, & \qquad $b_{12,12}=\varphi(t)\dfrac{n_{MG}}{N^\ast_M(t)}$. \medskip \\
\end{tabular}
$$

\clearpage

\addcontentsline{toc}{chapter}{

\end{document}